\documentclass[aps,prl,preprint,groupedaddress]{revtex4-1}
\usepackage{graphicx}
\usepackage{physics}

\usepackage{dcolumn}
\usepackage{bm}
\usepackage{amsmath}
\usepackage{subfigure}
\usepackage[none]{hyphenat}
 \usepackage{color}
 \usepackage{mathrsfs}

\usepackage{amssymb}%
\usepackage{mathrsfs}%

\begin{document}


\title{CETASim: A numerical tool for beam collective effect study in storage rings} 

\author{Chao Li}
\altaffiliation[]{li.chao@desy.de, yong-chul.chae@desy.de}
\author{Yong-Chul Chae}
\altaffiliation[]{yong-chul.chae@desy.de}
\affiliation{Deutsches Elektronen Synchrotron, Notkestrasse 85, 22607, Hamburg, Germany}

\begin{abstract}
We developed a 6D multi-particle tracking program CETASim in C++ programming language to simulate intensity-dependent effects in electron storage rings. The program can simulate the beam collective effects due to short-range/long-range wakefields for single/coupled-bunch instability studies. It also features to simulate interactions among charged ions and the trains of electron bunches, including both fast ion and ion trapping effects. The bunch-by-bunch feedback is also included so that the user can simulate the damping of the unstable motion when its growth rate is faster than the radiation damping rate. The particle dynamics is based on the one-turn map, including the nonlinear effects of amplitude-dependent tune shift, high-order chromaticity, and second-order momentum compaction factor. A skew quadrupole can also be introduced by the users, which is very useful for the emittance sharing and the emittance exchange studies. This paper describes the code structure, the physics models, and the algorithms used in CETASim. We also present the results of its application to PETRA-IV storage ring.

\begin{description}
\item[PACS numbers] 41.75.-i, 29.27.Bd, 29.20.Ej \\
\textbf{Key words:} {electron storage rings, beam collective effect, CETASim}
\end{description}
\end{abstract}

\maketitle

\section{1 Introduction}
The $4^{th}$ generation light sources move towards a diffraction-limited storage ring (DLSR) where the intense bunched beam with ultra-small emittance is stored for many hours for X-ray user operations.  Because of the small beam dimensions with appreciable beam intensities, various collective effects will limit the performance of the ring in delivering the optimum beam parameters for user operations. Traditionally, the short and long-range wakefield effects are the leading cause of instability we need to mitigate; however, we found that the ion, beam loading, transverse coupling, and other effects will also impact the beam parameters significantly. Since the Touschek lifetime impacts a whole aspect of operation, predicting and improving the lifetime becomes vital. These require investigating advanced beam dynamics caused by collective effects. 
     
Multi-particle tracking has been a popular method to investigate the collective effects in the electron storage rings; various codes have been developed including ELEGANT \cite{elegantWebPage}, MBTRACK \cite{skripka2016simultaneous}, and PyHEADTAIL \cite{PyHEADTAIL} $etc$. The benchmark between the codes is an ongoing effort in the light source communities. The motivation for developing CETASIm is to have a light and user-friendly tool, which includes the fundamental physics of various collective effects and approaches for instability mitigation. It is also beneficial for future studies since CETASim can be updated and upgraded appropriately when new physics needs arise. The remainder of this paper is organized as follows. In Section 2, the architecture of the code is introduced. Section 3 explains the physics models and algorithms for various collective effects used in CETASim. Section 4 showed simulation results by applying PETRA-IV storage ring as a demonstration. Summary and conclusions are given in the end

\section{2 General code overview}
\begin{figure}[!htp]
\centering
\includegraphics[width=1\textwidth]{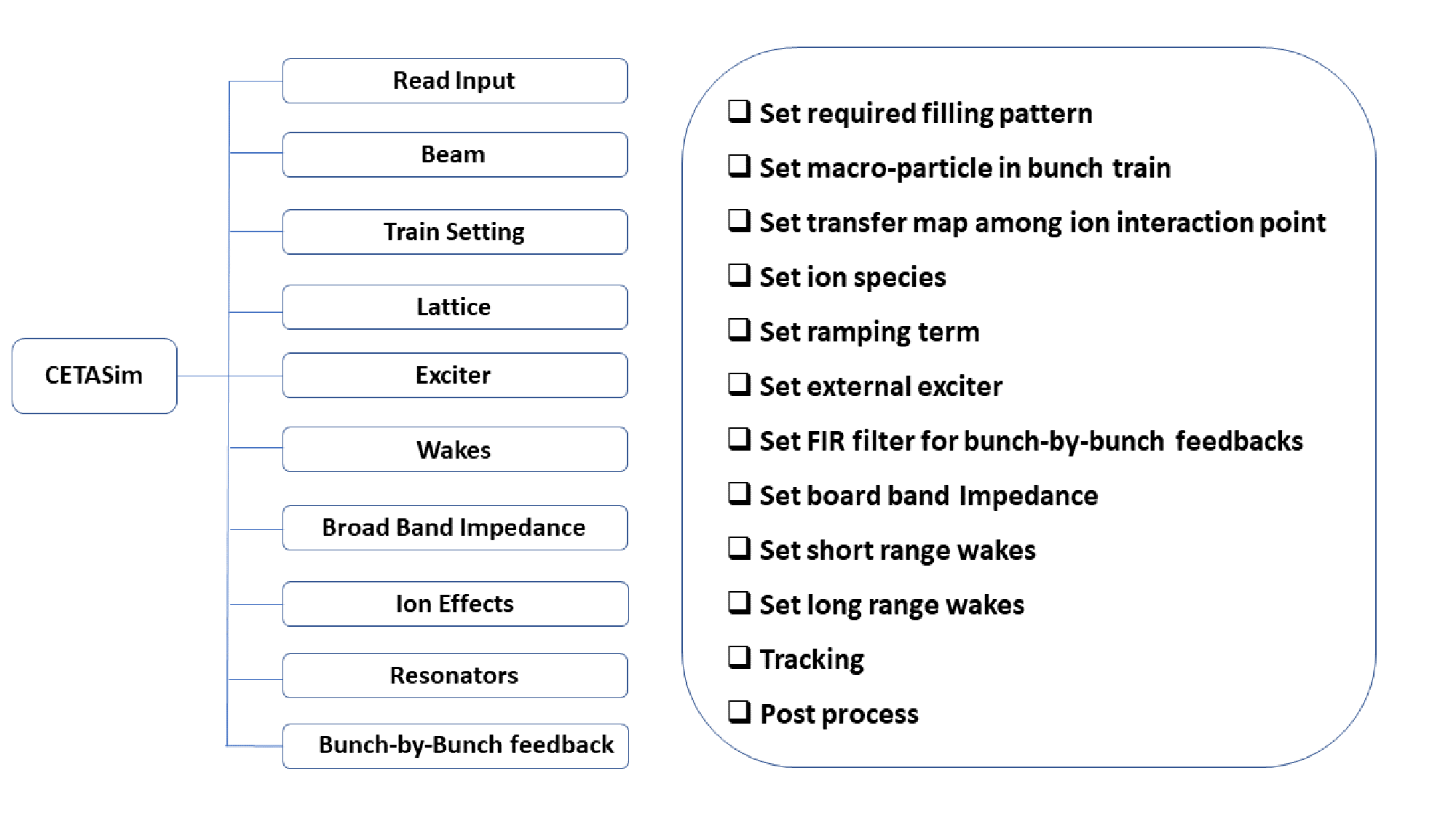}
\caption{General overview of CETASim code.}
\label{fig:2.1} 
\end{figure}

We developed CETASim following object-oriented concepts in C++ programming language \cite{CETASim}. Fig.~\ref{fig:2.1} shows the main classes designed in CETASim. The program generates the specified bunch train and launches the particle tracking task using the input parameters in the run setup. For the multi-bunch problem, the users can set each bunch charge individually while keeping the number of macro-particles per bunch constant. It is a handy feature for investigating transient beam loading compensation and ion cleaning when head and tail bunches are set to accommodate a larger bunch charge (guarding bunches). For ion effects investigation, CETASim can set multiple beam-ion interactions in the ring. Numerous ion species, local gas temperatures, and pressures can be set independently at each interaction point. The impedance class uses analytical formulas to construct the model impedance in a ring. At the current stage, two types of impedance elements, resistive wall and resonator model, are available in CETASim. Meanwhile, CETASim also can import impedance data from an external file. The exciter class can set an external exciter to the electron bunches with a given frequency, which plays as a coupled bunch mode driver in the "drive-damp" simulation. As a particular case of the longitudinal coupled bunch motion, the transient beam loading effect is dealt with by the resonator class. The cavity dynamics, which is driven by the generator current and the beam current, together with the beam dynamics are simulated simultaneously in a self-consistent manner.  We plan to implement the cavity feedback to stabilize the cavity voltage based on the control parameters in the future as well. The bunch-by-bunch feedback class adopts $Finite$ $Impulse$ $Response$ (FIR) filter to compute the kicker response based on the multi-turn beam position monitor (BPM) data. The data output uses the SDDS format, which can be post-processed by the SDDS toolkit \cite{SDDSWebpage}.

\section{3 Physical models in CETASim}
\subsection{3.1 Particle convention in CETASim}
In CETASim the position of every macro-particle is described by a 6D vector $\boldsymbol{x}=(x,p_x,y,p_y,dz,\delta)$ in phase space, where $x$ and $y$ are particle positions in the horizontal and vertical planes, $p_x $ and $ p_y$ are the horizontal and vertical momentum normalized by the reference momentum $p_0$, $dz$ and $\delta=(p-p_0)/p_0$ are the longitudinal positions and momentum deviation respecting to the reference particle. The sign convention is such that the head of the particles has a positive distance, namely, $dz > 0$. If time-dependent elements are included, such as RF cavities, $dz$ is converted to the time deviation by $d{\tau}=-dz/ \beta c$ to compute the RF phase, where $c$ is the speed of light, $\beta$ is the ratio of particle nominal velocity to $c$. 

\subsection{3.2 Beam transfer in one turn}
\subsubsection{3.2.1 The longitudinal beam transformation}
The longitudinal dynamics is described by 
\begin{equation}\label{eq:3.1.1}
dz_{i+1} = dz_{i} - L \sum_{j=1}^{3} \alpha_{cj} (\delta_{i})^{j}  \quad \quad
d\delta_{i+1} = d\delta_{i} + \frac{1}{\beta^2 E_0}(-U_0 + \sum_{n} e V_{n,rf} \sin(\omega_{n,rf} d\tau_{i}+\phi_n),   
\end{equation}
where $\alpha_{cj}$ is the $j$th order of momentum compaction factor, $E_0$ is the nominal particle total energy, $U_0$ is the energy loss per turn due to the synchrotron radiation, $n$ is the cavity index, $V_{n,rf}$, $\omega_{n,rf}$ and $\phi_n$ are the cavity voltage, angular frequency and phase respectively and $e$ is the electron charge. 

\subsubsection{3.2.2 Transverse plane }
The one-turn matrix in the transverse plane $\boldsymbol R$ is
\begin{equation}
\label{eq:3.1.2}
\boldsymbol R=
\left(\begin{array}{cccccc}
\cos\psi_x + \alpha_x \sin \psi_x & \beta_x \sin \psi_x  & 0 & 0 & 0 & R_{16}  \\ 
-\gamma_x \sin\psi_x  & \cos\psi_x - \alpha_x \sin \psi_x & 0 & 0 & 0 & R_{26} \\ 
0 & 0 & \cos\psi_y + \alpha_y \sin \psi_y & \beta_y \sin \psi_y  & 0  & R_{36} \\ 
0 & 0 & -\gamma_y \sin\psi_y  & \cos\psi_y - \alpha_y \sin \psi_y & 0 & R_{46} \\ 
R_{51} & R_{52} & R_{53} & R_{54}  & 1  & 0   \\ 
0 & 0 & 0 & 0  & 0  & 1   
\end{array}\right) \\
\end{equation}
where $\alpha_{(x,y)}$, $\beta_{(x,y)}$ and $\gamma_{(x,y)}$ are the Twiss parameters. $R_{5j}$ and $R_{i6}$ are dispersion-related functions, which ensure the closed orbit condition. Noticeably, $R_{55} = R_{66} = 1$.  
$\psi_{(x,y)}$ is the phase advance per turn, including the contribution from chromaticity and amplitude-dependent tune shift \cite{elegantWebPage}. Explicitly, the phase advance $\psi_{(x,y)}=2\pi \nu_{(x,y)}$. The tunes applied in tracking are 
\begin{equation}
\label{eq:3.0.0}
\begin{split}
\nu_x &= \nu_{x,0} + \xi_x \delta + ADT_x^1 A_x + \frac{ADT_x^2}{2} A_x^2 + ADT_{xy}^2 A_x A_y \\
\nu_y &= \nu_{y,0} + \xi_y \delta + ADT_y^1 A_y + \frac{ADT_y^2}{2} A_y^2 + ADT_{yx}^2 A_y A_x
\end{split}
\end{equation}
where $\nu_{(x,y),0}$ is the nominal tune, $\xi_{(x,y)}$ is the chromaticity, $ADT^{i}$ is the amplitude dependent tune coefficient. $A_x$ and $A_y$ are the single particle oscillation amplitude in the horizontal and vertical planes 
\begin{equation}\label{eq:3.0.1}
A_x = \frac{x_{\beta}^2 + (\alpha_x x_{\beta} + \beta_x x_{\beta}')^2}{\beta_x} \quad \quad \quad
A_y = \frac{y_{\beta}^2 + (\alpha_y y_{\beta} + \beta_y y_{\beta}')^2}{\beta_x}, 
\end{equation}
where ($x_{\beta}$ $x_{\beta}'$) and ($y_{\beta}$ $y_{\beta}'$) are the particle betatron oscillation excluding the influence due to dispersion. 

For the beam-ion interaction in CETASim, users can split the ring into several sections to set up multiple beam-ion interaction points. In that case, a simpler map representing beam transfer from the $i$th to $(i+1)$th beam-ion interaction point is adopted for the transverse plane
\begin{equation}
\label{eq:3.1.3}
\boldsymbol R_x^{i+1|i}=
\left(\begin{array}{ccc}
\sqrt{\frac{\beta_{x,i+1}}{\beta_{x,i}}}(\cos\psi_{xi} + \alpha_{x,i} \sin \psi_{xi}) & \sqrt{\beta_{x,i+1}\beta_{x,i}} \sin \psi_{xi}  & \eta_{x,i}  \\ 
-\frac{1+\alpha_{x,i+1} \alpha_{x,i}}{\sqrt{\beta_{x,i+1} \beta_{x,i}}} \sin\psi_{xi} +\frac{\alpha_{x,i} - \alpha_{x,i+1}}{\sqrt{\beta_{x,i+1} \beta_{x,i}}} \cos\psi_{xi}   & \sqrt{\frac{\beta_{x,i}}{\beta_{x,i+1}}}(\cos\psi_{xi} - \alpha_{x,i+1} \sin \psi_{xi}) & \eta'_{xi} \\ 
0 & 0 &  1 \\ 
\end{array}\right)
\end{equation}
where $\alpha_{xi}$, $\beta_{xi}$, $\psi_{xi}$ are the local Twiss parameters and  phase advance, $\eta_{xi}$, $\eta'_{xi}$ are the dispersion functions. The transfer map $\boldsymbol R_y^{i+1|i}$ in the vertical plane is obtained similarly.

\subsubsection{3.2.3 synchrotron radiation and quantum excitation}
We follow the approaches in BBC code \cite{hirata1997bbc} from Hirata to simulate the synchrotron radiation and quantum excitation effects. To do that, we normalize the accelerator coordinate vector $\boldsymbol{x}$ to normal frame $\boldsymbol{X}$  by using dispersion matrix  $\boldsymbol{H}$ and Twiss matrix $\boldsymbol{B}$
\begin{equation}
\label{eq:3.1.4}
\boldsymbol{X} =  \boldsymbol{B} \boldsymbol{H} \boldsymbol{x}. 
\end{equation}
The dispersion matrix $\boldsymbol{H}$ is
\begin{equation}
\label{eq:3.1.7}
\boldsymbol H = 
\left(\begin{array}{ccc}
\boldsymbol I & 0          & -\boldsymbol H_x  \\ 
0         & \boldsymbol I  & -\boldsymbol H_y \\ 
-\boldsymbol {J_2 H_x^T J_2} & -\boldsymbol {J_2 H_y^T J_2} &  \boldsymbol I \\ 
\end{array}\right), 
\end{equation}
where 
\begin{equation}
\label{eq:3.1.8}
\boldsymbol H_x=
\left(\begin{array}{cc}
0        &  \eta_{x}      \\ 
0         & \eta'_{x}       \\  
\end{array}\right), \quad
\boldsymbol H_y=
\left(\begin{array}{cc}
0        &  \eta_{y}      \\ 
0         & \eta'_{y}       \\  
\end{array}\right), \quad
\boldsymbol J_2=
\left(\begin{array}{cc}
0        &  -1      \\ 
-1       &  0       \\  
\end{array}\right).
\end{equation}
The Tiwss matrix $\boldsymbol{B}$ is 
\begin{equation}
\label{eq:3.1.9}
\boldsymbol B=
\left(\begin{array}{cccccc}
\frac{1}{\sqrt{\beta_x}}         & 0 & 0 & 0 & 0 & 0 \\ 
\frac{\alpha_x}{\sqrt{\beta_x}}  & \sqrt{\beta_x} & 0 & 0 & 0 & 0 \\ 
0 & 0 & \frac{1}{\sqrt{\beta_y}} & 0  & 0  & 0 \\ 
0 & 0 & \frac{\alpha_y}{\sqrt{\beta_y}}  & \sqrt{\beta_y} & 0 & 0 \\ 
0 & 0 & 0 & 0  & \frac{1}{\sqrt{\beta_z}}  & 0   \\ 
0 & 0 & 0 & 0  & \frac{\alpha_z}{\sqrt{\beta_z}}  & \sqrt{\beta_z}
\end{array}\right).
\end{equation}

In CETASim, the Twiss parameter $\alpha_z$ is assumed to be zero, $\beta_z=\sigma_z/\sigma_e$, where $\sigma_z$ and $\sigma_e$ are the natural bunch length and energy spread. Thereafter, the effect of synchrotron radiation and quantum excitation are simulated in the normalized frame $\boldsymbol{X}$ once per turn according to 
\begin{equation}
\label{eq:3.1.5}
\begin{split}
\left(\begin{array}{c}
\boldsymbol{X_1} \\ \boldsymbol{X_2}  
\end{array}\right)
&= \lambda_x
\left(\begin{array}{c}
\boldsymbol{X_1} \\ \boldsymbol{X_2}
\end{array}\right)
+ \sqrt{\epsilon_x(1-\lambda_x^2)}
\left(\begin{array}{c}
\hat r_1 \\ \hat r_2  
\end{array}\right) \\
\left(\begin{array}{c}
\boldsymbol{X_3} \\ \boldsymbol{X_4}  
\end{array}\right)
&= \lambda_y
\left(\begin{array}{c}
\boldsymbol{X_3} \\ \boldsymbol{X_4}
\end{array}\right)
+ \sqrt{\epsilon_y(1-\lambda_y^2)}
\left(\begin{array}{c}
\hat r_3 \\ \hat r_4  
\end{array}\right) \\
\left(\begin{array}{c}
\boldsymbol{X_5} \\ \boldsymbol{X_6}  
\end{array}\right)
&= 
\left(\begin{array}{cc}
1 & 0  \\ 
0 & \lambda_z^2
\end{array}\right)
\left(\begin{array}{c}
\boldsymbol{X_5} \\ \boldsymbol{X_6}
\end{array}\right)
+
\left(\begin{array}{c}
0  \\  \sqrt{\epsilon_z(1-\lambda_z^4)}
\end{array}\right)
\left(\begin{array}{c}
0 \\ \hat r_6  
\end{array}\right).
\end{split}
\end{equation}
Here $\hat r$ is an independent Gaussian random variable with a unit variance; $\lambda_{x,y,z}=\exp(-1/\tau_{x,y,z})$ is the transport coefficient with $\tau_{x,y,z}$ representing the synchrotron radiation damping time in the unit of the number of turns; $\epsilon_x$,  $\epsilon_y$ and $\epsilon_z$ are the equilibrium beam emittance, where $\epsilon_z=\sigma_z\sigma_e$. Once the synchrotron radiation and quantum excitation simulation is done, the normalized $\boldsymbol X$ is transformed back to  $\boldsymbol x$ by
\begin{equation}
\label{eq:3.1.6}
\boldsymbol{x} =  \boldsymbol{H}^{-1} \boldsymbol{B}^{-1} \boldsymbol{X}.
\end{equation}

\subsection{3.3 Impedance and wakes models in CETASim}
In CETASim, the single-bunch effect is simulated by using the concept of impedance~\cite{AlexChao, panofsky1956some, vaganian1995panofsky} in the frequency domain; the coupled bunch effect is simulated in the time domain by using the analytical formula of resistive wall and resonator wakes~\cite{AlexChao}. The board band impedance data for the single bunch effect study can be generated from analytical models or imported from an external file. 

\subsubsection{3.3.1 RLC impedance and wake}
Equation~\ref{eq:3.2.1} gives the impedance of the RLC circuit as a function of angular frequency $\omega$, 
\begin{equation}\label{eq:3.2.1}
Z_m^{\parallel}(\omega) = \frac{\omega}{c} Z_m^{\perp}(\omega) = \frac{R_s}{1+iQ(\frac{\omega_r}{\omega}-\frac{\omega}{\omega_r})},  
\end{equation}
where $R_s$ is the resistance with a dimension $\Omega/L^{2m}$, $Q$ is the quality factor, $\omega_r$ is the resonant angular frequency.
Correspondingly, the longitudinal ($W_m'(z)$) and the transverse ($W_m(z)$) wake functions are
\begin{equation}\label{eq:3.2.2}
\begin{split} 
W_m'(z) &=  \left\{  \begin{array}{cc}
0              		   																																& z>0     \\
\frac{R_s}{\tau_f}     																															    & z=0     \\
2 \frac{R_s}{\tau_f} e^{ z /c \tau_f} (\cos \frac{\overline{\omega}z}{c}  + \frac{1}{\overline{\omega} \tau_f} \sin \frac{\overline{\omega}z}{c})   & z<0,
\end{array} \right.  \\
W_m(z) &= \left \{ \begin{array}{cc}
0       																		 & z=0  \\
\frac{ c R_s \omega_r}{2 Q} e^{z /c \tau_f}  \sin \frac{\overline{\omega}z}{c}   & z<=0,
\end{array} \right. \\
\end{split}
\end{equation}
where $\tau_f=2Q/\omega_r$ represents the filling or damping time in the circuit, $\overline{\omega}=\sqrt{\omega_r^2 - 1/\tau_f^2}$.

\subsubsection{3.3.2 Resistive wall impedance and wake }
The resistive wall (RW) impedance from an infinitely thick metallic round beam pipe of radius $b$ and conductivity $\sigma$ is well-known \cite{AlexChao}. When the beam pipe has an elliptical shape, Yokoya form factors \cite{yokoya1993resistive} are applied for correction. Eq.~\ref{eq:3.2.3} shows the model of RW impedance per unit length as a function of angular frequency $\omega$ 
\begin{equation}\label{eq:3.2.3}
\begin{split}
\frac{Z_m^{\parallel}(\omega)}{L} = \frac{\omega}{c} \frac{Z_m^{\perp}(\omega)}{L} &= 
\frac{4/b^{2m}}{(1+\delta_{m0}) b c \sqrt{\frac{2\pi\sigma}{\abs{\omega}}}
[1+sgn(\omega)i]-\frac{ib^2}{m+1}\omega + \frac{imc^2}{\omega}} \\
&\approx \sqrt{\frac{2}{\pi \sigma}} \frac{1}{(1+\delta_{m0})b^{2m+1}c}\abs{\omega}[1-sgn(\omega)i], 
\end{split}
\end{equation}
where $\delta_{m,0}$ is the Kronecker-Delta function, $i$ is the imaginary unit, $sgn$ is the $Sign$ function. Limiting the RW impedance to the lowest order $m=0,1$ in the longitudinal and transverse respectively, the longitudinal and transverse wake functions can be obtained by Fourier transformation. Define $z_0=(2\chi)^{1/3}b$ as the characteristic distance, where $\chi=c/(4 \pi \sigma b)$ is a dimensionless parameter, the longitudinal and transverse RW wakes are \cite{skripka2016simultaneous}:
\begin{equation}\label{eq:3.2.4}
\begin{split}
W_0'(z) &= \frac{1}{b^2} [ \frac{e^{-z/z_0}}{3} \cos(\frac{ \sqrt{3} z}{z_0}) - \frac{\sqrt{2}}{\pi} \int_0^{\infty}dx \frac{x^2 e^{-x^2 z/z_0}}{x^6+8}] \\
W_1(z)  &= -\frac{32}{b^3} (2 \chi)^{1/3} [\frac{e^{-z/z_0}}{12} \cos(\frac{ \sqrt{3} z}{z_0}) - \frac{1}{4 \sqrt{3}} e^{-z/z_0} \sin(\frac{ \sqrt{3} z}{z_0}) - \frac{\sqrt{2}}{\pi} \int_0^{\infty}dx \frac{x^2 e^{-x^2 z/z_0}}{x^6+8}]. 
\end{split}
\end{equation}
If the simplified RW impedance, Eq.~\ref{eq:3.2.3}, is adopted, the RW wakes are simplified further to  
\begin{equation}
\label{eq:3.2.5}
W_0'(z) \approx \frac{1}{2\pi b }\sqrt{\frac{c}{\sigma}} \frac{1}{|z|^{3/2}} \quad \quad
W_1(z)  \approx -\frac{2}{\pi b^3}  \sqrt{\frac{c}{\sigma}}  \frac{1}{|z|^{1/2}}.
\end{equation}

\begin{figure}[!htp]
\centering
\subfigure{
    \label{fig3.1_a} 
    \includegraphics[width=2.5in]{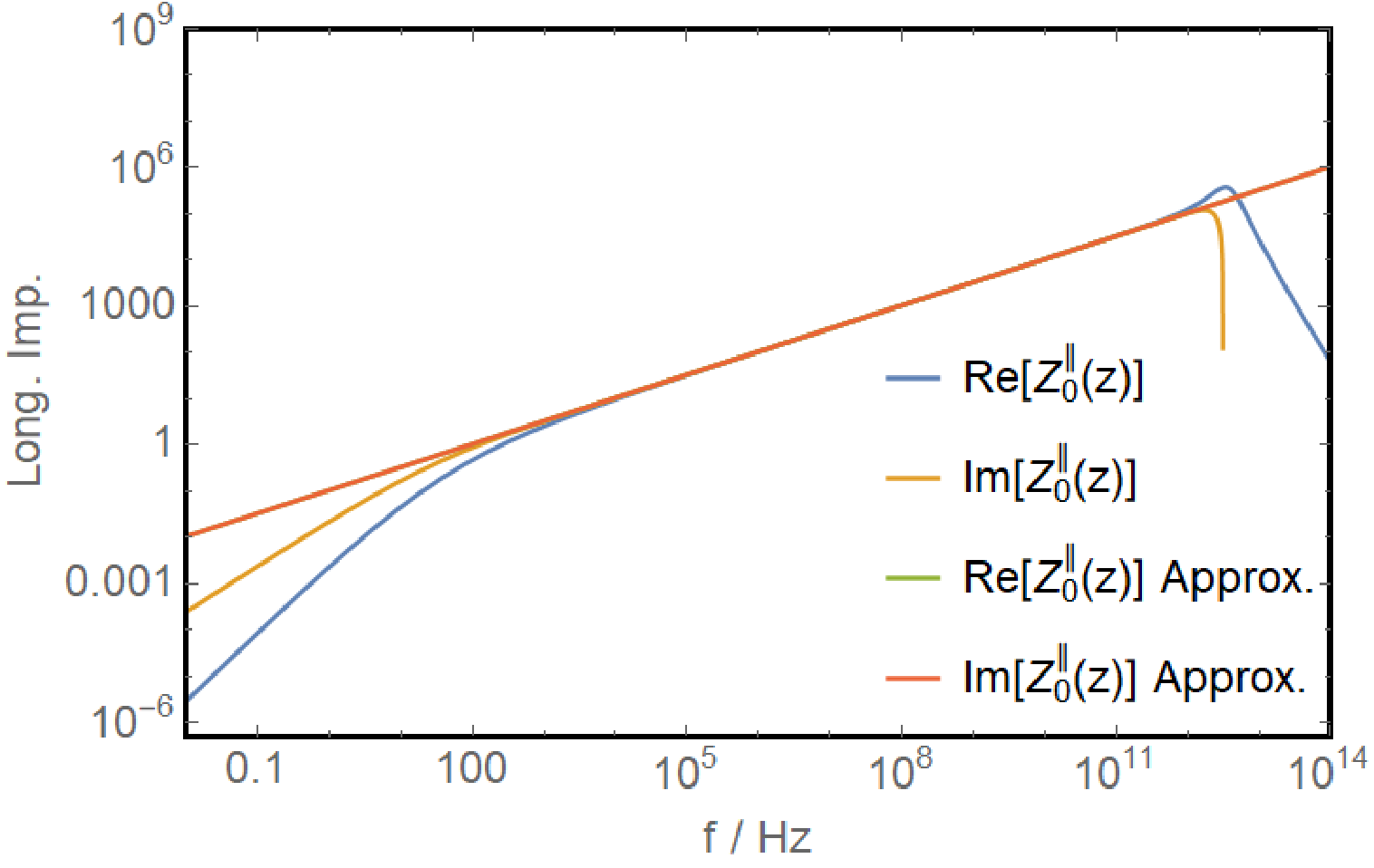}}
\subfigure{
    \label{fig3.1_b} 
    \includegraphics[width=2.5in]{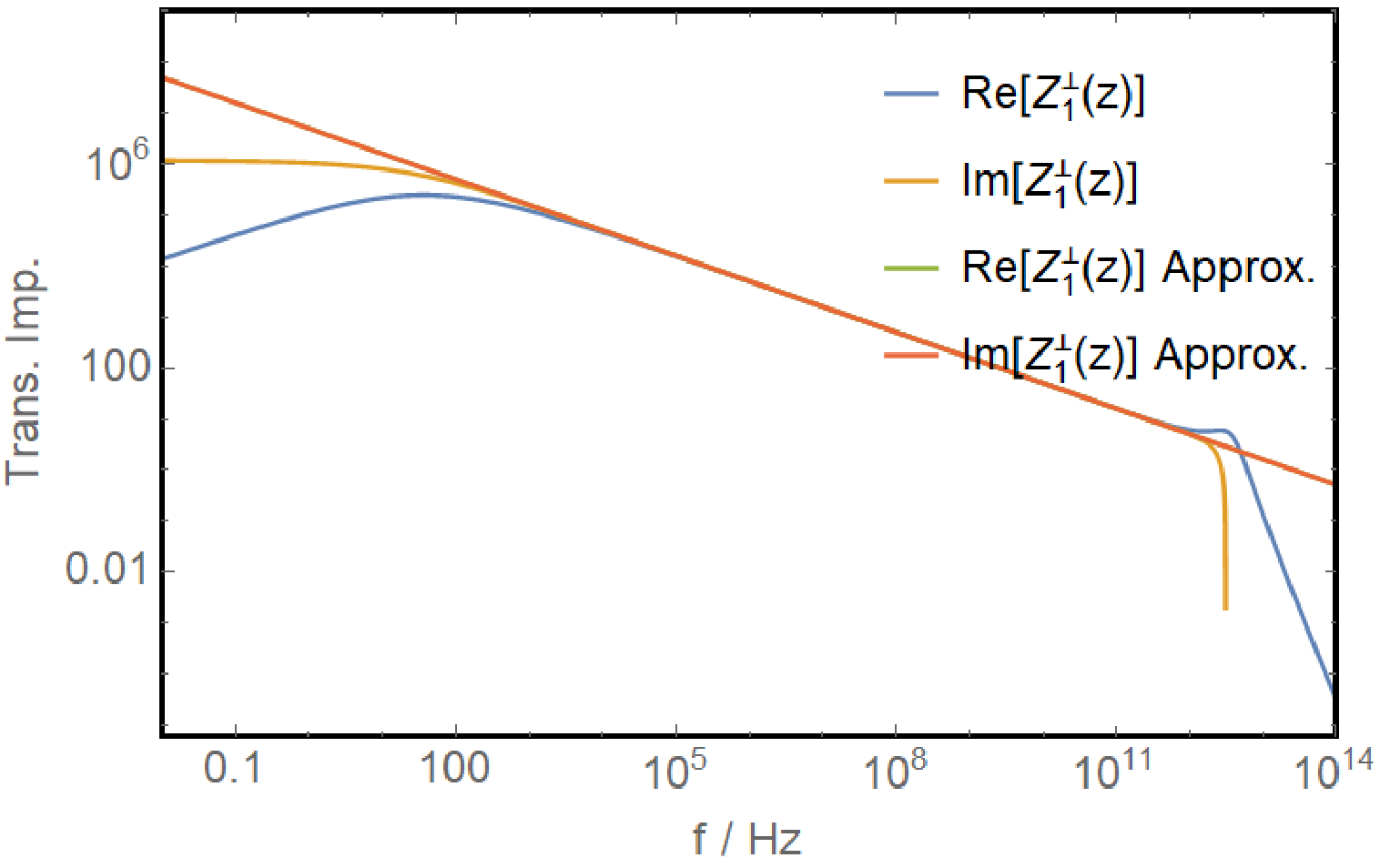}}
\subfigure{
    \label{fig3.1_c} 
    \includegraphics[width=2.5in]{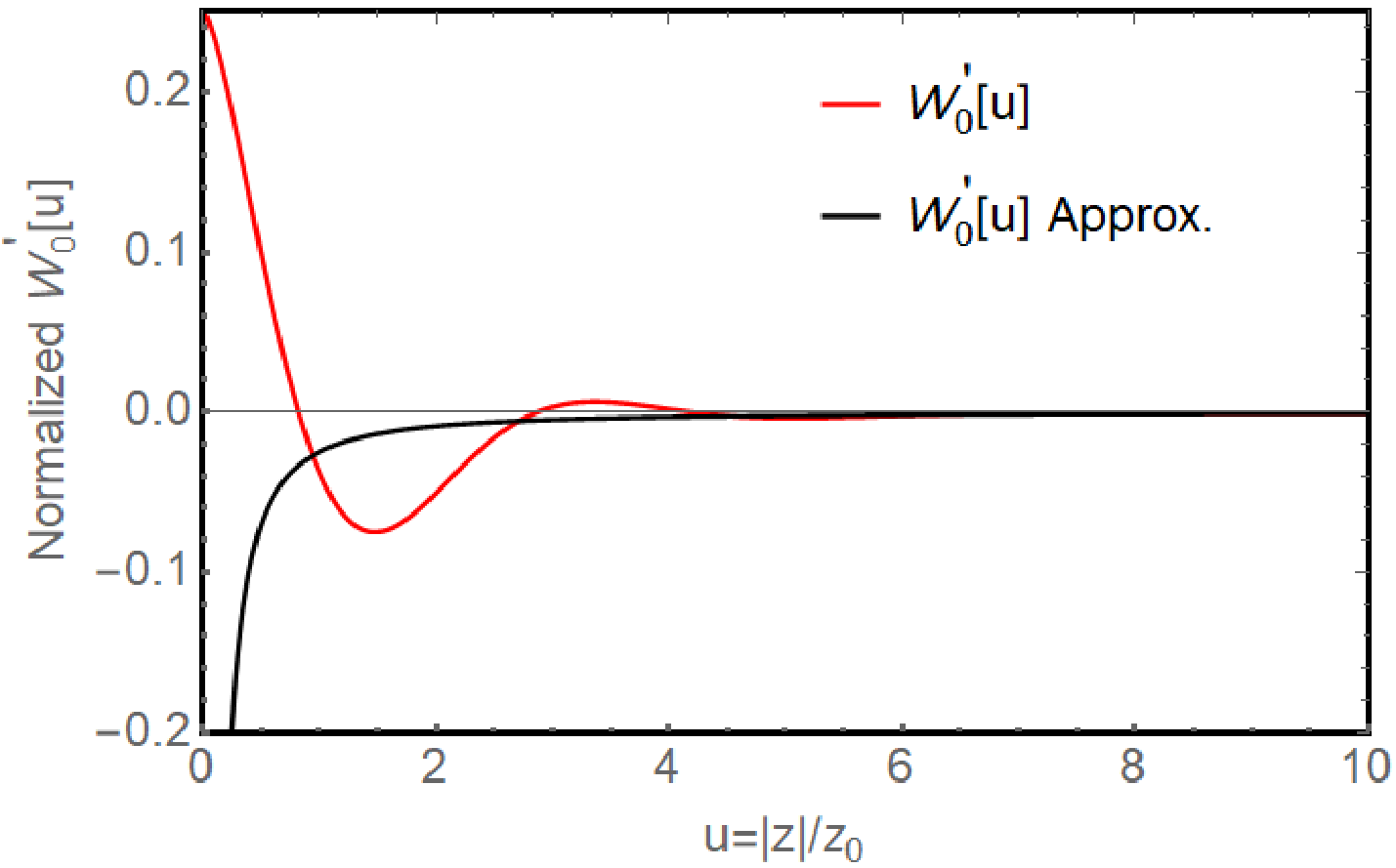}}
\subfigure{
    \label{fig3.1_d} 
    \includegraphics[width=2.5in]{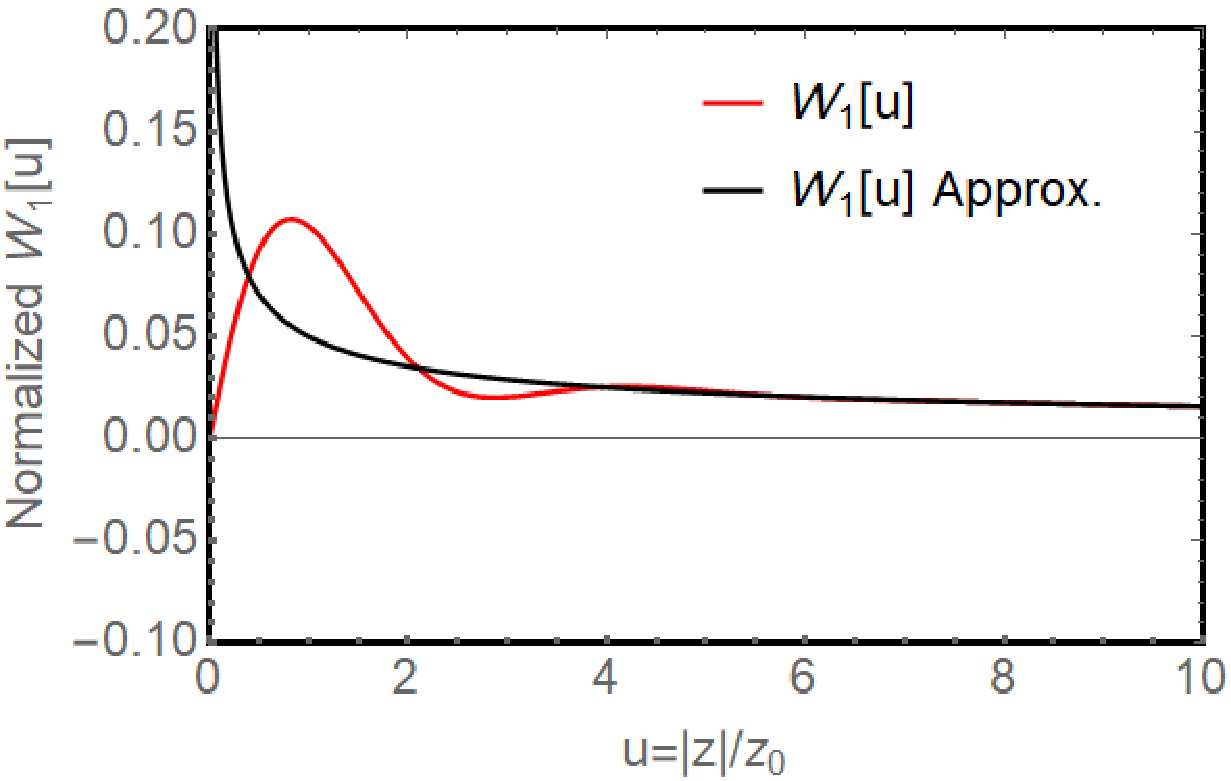}}    
\caption{\label{fig:3.1} Comparison of the exact and approximated solutions of the resistive impedance (above) and wakes (bottom). The sub-figures on the left and right correspond to the longitudinal ($m=0$) and the transverse ($m=1$) respectively.}
\end{figure}

Figure~\ref{fig:3.1} compares the exact and approximate solution of the RW impedance and wakes. In terms of impedance, the discrepancy exists both in the low and high-frequency regions. The results from the exact and the approximated solutions are the same in the medium frequency region. The wakes from the exact and approximate solutions agree well when $u=z/z_0>5$. Since the distance among different bunches is much longer than the characteristic distance $z_0$, the asymptotic wake Eq.~\ref{eq:3.2.5} is applied in CETASim for the coupled-bunched simulation to relax the computation load.

\subsection{3.4 Single bunch effect}
The energy change within the bunch caused by the longitudinal wake or impedance can be expressed as
\begin{equation}\label{eq:3.3.1}
eV^{\parallel}(z) = -e\int_z^{\infty} \rho(z') W_0'(z-z')dz' =-\frac{e}{2\pi} \int_{-\infty}^{\infty} e^{i \omega z /c} \widetilde{\rho}(\omega) Z_0^{\parallel}(\omega) d\omega,
\end{equation}
where $\widetilde{\rho}(\omega)$ is the charge-density spectrum in frequency domain. In the transverse plane, we consider the dipole $W^D_1$ and quadrupole $W^Q_1$ wakes, the impedance of which are denoted as $Z^D_1(\omega)$ and $Z^Q_1(\omega)$ respectively. The 'dipole' and 'quadrupole' represent that the kick particle experiences are proportional to the position offset of the previous particle and its position offset respectively\footnote{The quadrupole impedance $Z_1^{Q}(\omega)$ and quadrupole wake $W^Q_1$ discussed in this paper are also termed as de-tuning impedance and de-tuning transverse wakes.} \cite{Heifets:1998er, Mounet:1451296}. The transverse kick from the impedance within the bunch can be expressed as
\begin{equation}\label{eq:3.3.2}
\begin{split} 
eV^{\perp}(z) &=  -e\int_z^{\infty} \rho(z') x(z') W^D_1(z-z')dz' -e\int_z^{\infty} \rho(z') x(z) W^Q_1(z-z')dz' \\
&=-i\frac{e}{2\pi} \int_{-\infty}^{\infty} e^{i \omega z /c} \tilde{\rho}^D(\omega) Z_1^{D}(\omega) d\omega - i\frac{e}{2\pi}x(z) \int_{-\infty}^{\infty} e^{i \omega z /c} \widetilde{\rho}(\omega) Z_1^{Q}(\omega) d\omega,
\end{split}
\end{equation}
where, $x(z)$ is the transverse position offset along the beam, $\widetilde{\rho}^D(\omega)$ is the the spectrum of the dipole moment $\rho(z)x(z)$. 

For a given impedance, Eq.~\ref{eq:3.3.1} and Eq.~\ref{eq:3.3.2} in frequency domain, are used to compute $eV^{\parallel}(z)$ and $eV^{\perp}(z)$ in CETASim. In the future, if there is a requirement to do the simulation with the time domain method, a subroutine can be developed to export the wakes from an external file supplied by the user to get $eV^{\parallel}(z)$ and $eV^{\perp}(z)$ as well. 

\subsection{3.5 Coupled bunch effect}
If the wakes are generated by the RW or high-Q resonators which last longer than the bunch spacing, the motion of bunches will be coupled. Assuming there are $N_b$ bunches in total in the ring, the kick at the $j$th bunch at the $n$th turn from the longitudinal wakes $W'_0$ and transverse dipole ($W_1^D$) and quadrupole wakes ($W_1^Q$) can be expressed as
\begin{equation}\label{eq:3.4.1}
\begin{split} 
\Delta p^{j}_{z;n} &\sim \sum_{k=0}^{n-1}\sum_{i=0}^{Nb-1} W'_0( h c T_{rf} (n-k) + \delta s_{i;j})  \\ 
\Delta p^{j}_{x;n} &\sim \sum_{k=0}^{n-1}\sum_{i=0}^{Nb-1} [\langle x \rangle^i_k W_1^D( h c T_{rf} (n-k)  + \delta s_{i;j}) + \langle x \rangle^j_{0} W_1^Q( h c T_{rf} (n-k)  + \delta s_{i;j}) ], \\ 
\delta s_{i;j} &= 
\left\{ \begin{array}{cc}
(i-j) c T_{rf}              & j<=i     \\
(i-j+h) c T_{rf}            & j>i     
\end{array} \right. \\
\end{split}
\end{equation}
where $h$ is the RF harmonic number, $T_{rf}$ is the RF period, $\langle x \rangle^i_k$ is bunch centroid  of the $i$th bunch at $k$th previous turn. CETASim can simulate the long-range wakes from the RW (Eq.~\ref{eq:3.2.2}) and RLC (Eq.~\ref{eq:3.2.5}) elements. The number of turns the wakes last can be specified by the users.

\subsection{3.6 Bunch-by-bunch feedback}
The bunch-by-bunch feedback can mitigate the coupled bunch instabilities. It detects the transverse or longitudinal positions and creates the kicker signal in the bunch-by-bunch sense. A digital FIR filter to process the position signal to the kicker is expressed as
\begin{equation}\label{eq:3.5.1}
\Theta_n = \sum_{k=0}^N a_k \theta_{n-k}
\end{equation}
where $a_k$ represents the filter coefficient, $\theta_{n-k}$ and $\Theta_n$ are the input and output of the filter. The number of the input data $N+1$ is defined as taps of the filter. With a given coefficient, the bunch momentum change at the kicker in the $n$th turn due to the bunch-by-bunch feedback can be found  
\begin{equation}\label{eq:3.5.2}
\Delta p_{x;n} \sim K_x \sum_{k=0}^{N} a_{k,x} \langle x \rangle_{n-k}, \quad
\Delta p_{y;n} \sim K_y \sum_{k=0}^{N} a_{k,y} \langle y \rangle_{n-k}, \quad
\Delta \delta_{n} \sim K_z \sum_{k=0}^{N} a_{k,z} \langle z \rangle_{n-k},
\end{equation}
where $\langle x,y,z \rangle_{n-k}$ is the bunch centroids of the $k$th previous turn at the BPM, $K_{x,y,z}$ is the gain factor. Particles in one bunch experience the same kick. The shunt impedance and power constraint of the kicker, specified by the users, limit the maximum kick strength experienced by particles.    
 
\subsection{3.7 Beam-ion effect}
CETASim simulates the Coulomb interaction between the electron bunch and the ionized gases~\cite{kohaupt1971mechanismus}. In accelerators, if the gas pressure $P$, gas temperature $T$ and ionization collision cross-section $\Sigma$ are given, the number of ionized gases can be obtained by
\begin{equation}\label{eq:3.6.1}
\Lambda =\Sigma \frac{P N_A}{R T} N_b, 
\end{equation}
where $N_b$ is the number of electron particles, $R$ the ideal gas constant and $N_A$ is the Avogadro constant. The ions can be locally trapped by passing electron bunches acting as the focusing lenses. The trapping condition is
\begin{equation}\label{eq:3.6.4}
A \geq A_{c}=\frac{Q N_b r_p \Delta T_b c}{2 \sigma_y (\sigma_x + \sigma_y) }
\end{equation}
where $A$ is the ion mass, $Q$ is the ion charge number, $r_p$ is the classical radius of the proton, $\Delta T_b$ is the bunch separation in time. Only the ions with a mass number larger than the critical mass number $A_{c}$ can be trapped. This formula assumes a linear focusing and uniform fill pattern. Note that the critical ion mass $A_{c}$ varies following the betatron functions around the ring.  

In CETASim, we assume the electron bunch has a rigid Gaussian distribution. The $Bassetti-Erskine$ formula \cite{Bassetti} is applied at the interaction point $s_i$ to get the beam ion interaction force
\begin{equation}\label{eq:3.6.3}
\begin{split}
\begin{aligned}
\boldsymbol{F_y(x,y)} + i \boldsymbol{F_x(x,y)} &= \delta(s_i) \sqrt{\frac{\pi}{2(\sigma_x^2 -\sigma_y^2)}}  \\
&\{w(\frac{x+iy}{ \sqrt{2\pi (\sigma_x^2 -\sigma_y^2)}}) 
-\exp(-\frac{x^2}{2\sigma_x^2}-\frac{y^2}{2\sigma_y^2})w(\frac{x\frac{\sigma_y}{\sigma_x}+iy\frac{\sigma_x}{\sigma_y} }{\sqrt{2\pi (\sigma_x^2 -\sigma_y^2)}})\},
\end{aligned}
\end{split}
\end{equation}
where  $w(z)$ is the complex error function, $\sigma_x$ and $\sigma_y$ are the rms bunch size \footnote{To avoid the numerical divergence, in CETASim, Eq.~\ref{eq:3.6.3} is de-generates to the 1D Gaussian model when  $abs((\sigma_x - \sigma_y)/\sigma_y)<10^{-4}$.}. Note $N_i$ as ionized ions, the motions of electrons and ions follow the equation
\begin{equation}\label{eq:3.6.2}
\begin{split}
\begin{aligned}
\frac{d^2 \boldsymbol{\langle x \rangle}}{ds^2} = \frac{2r_e}{\gamma}\sum_{i=0}^{N_i} \boldsymbol{F}(\boldsymbol{\langle x \rangle} - \boldsymbol{X_i}), \quad \quad
\frac{d^2 \boldsymbol{X_i}}{dt^2} = \frac{2 N_b r_e c^2}{M_i/m_e} \sum_{j=0}^{N_b} \boldsymbol{F}(\boldsymbol{X_i} - \boldsymbol{\langle x \rangle}),
\end{aligned}
\end{split}
\end{equation}
where $\boldsymbol{X_i}=(x,y)_i$ is the ion position, $\boldsymbol{\langle x \rangle}=({\langle x \rangle},{\langle y \rangle})$ is the bunch centroids, $\boldsymbol{F}$ is the Coulomb force between the ions and electron particles, $r_e$ is the classical electron radius, $\gamma$ is the relativistic factor of electron beam, $M_i$ and $m_e$ are the mass of ion and electron. This model assumes that the electron bunch is rigid (weak-strong model). If the motion of individual electrons is important, one can assume that ion distribution follows a Gaussian profile as well. Then the Coulomb force from the ions on individual electron particles can be obtained similarly (quasi-strong-strong model). In CETASim, both the weak-strong and quasi-strong-strong models are available.

\subsection{3.8 Beam loading effect and cavity feedback}
The longitudinal dynamics due to the RF mode in the cavity can be simulated as well. CSTASim follows the algorithm in P.B. Wilson’s paper \cite{wilson1994fundamental}, where the beam-induced voltage is treated in the phasor frame.  The force electron particle experiences in terms of beam-included voltage in the phasor fame is equivalent to the force in terms of the long-range wake. However, in the phasor frame, the history of the bunches in previous turns is not needed any longer, which reduces the computation load significantly.

The transient beam loading effect particularly refers to the fundamental cavity mode, which is the same mode building up the acceleration field. By modeling the fundamental mode of the cavity as an RLC circuit, the generator dynamics and beam dynamics are coupled. In a ring with multiple RF systems, at least the main cavity has to be active, which brings additional complexity to the whole study, especially when the cavity feedback has to be included further.

In CETASim, the total cavity voltage $\boldsymbol{\tilde{V}_c(t)}$, that beam can sample as a function of time $t$, is the sum of the generator voltage $\boldsymbol{\tilde{V}_{g}}(t)$ and beam-induced voltage $\boldsymbol{\tilde{V}_{b}}(t)$,
\begin{equation}
\label{eq:3.7.1}
\boldsymbol{\tilde{V}_c(t)} = \boldsymbol{\tilde{V}_{g}}(t) + \boldsymbol{\tilde{V}_{b}}(t).
\end{equation}
$\boldsymbol{\tilde{V}_{b}}(t)$ and $\boldsymbol{\tilde{V}_{g}}(t)$ are driven by the beam current $\boldsymbol{\tilde{I}_b}(t)$ and the generator current $\boldsymbol{\tilde{I}_{g}}(t)$ respectively. With the RLC circuit model, the resonant voltage excited  by a driving current $\boldsymbol{\tilde{I}(t)}$ follows the differential equation \cite{schilcher1998vector} 
\begin{equation}
\label{eq:3.7.2}
\frac{d^2}{dt^2}\boldsymbol{\tilde{V}}(t) + \frac{\omega_{r}}{Q_{L}} \frac{d}{dt}\boldsymbol{\tilde{V}}(t) + \omega_{r}^2 \boldsymbol{\tilde{V}}(t) = \frac{\omega_{r} R_{L}}{Q_L} \frac{d}{dt}\boldsymbol{\tilde{I}}(t),
\end{equation}
where  $Q_L$ and $R_L$ are the loaded quality factor and shunt impedance.

Note $\omega_{rf}$ as the angular frequency of the driving term $\boldsymbol{\tilde{I}}(t)$, if several conditions are met: (1) the second-order terms can be neglected; (2) $Q_L\gg1$; (3) $\Delta\omega=\omega_r-\omega_{rf}\ll\omega_r$, and together with the zero-order hold method, the solution of Eq.~\ref{eq:3.7.2} can be further simplified in the state space \cite{schilcher1998vector,berenc2015modeling}    
\begin{equation}
\label{eq:3.7.3}
\begin{aligned}
\left(\begin{array}{c} 
V^r  \\ 
V^i   
\end{array}\right)_{t+\Delta t}
&= e^{-\omega_{1/2} \Delta t}
\left(\begin{array}{cc} 
\cos \Delta \omega \Delta t  & -\sin \Delta \omega \Delta t     \\ 
\sin \Delta \omega \Delta t  &  \cos \Delta \omega \Delta t 
\end{array}\right)
\left(\begin{array}{c}
V^r  \\ 
V^i  
\end{array}\right)  \\
 &+ \frac{\omega_r R_L}{2 Q_L (\omega_{1/2}^2 + \Delta \omega^2) }
\left(\begin{array}{cc}
A  & B   \\ 
-B   &  A
\end{array}\right)
\left(\begin{array}{c}
I^r  \\ 
I^i   \\ 
\end{array}\right)_t \\
\end{aligned}
\end{equation}
where $\omega_{1/2}=1/\tau_f$, $\Delta t$ is the time step, $A$ and $B$ are
\begin{equation}
\label{eq:3.7.4}
\begin{aligned}
A &= \Delta \omega e^{-\omega_{1/2} \Delta t} \sin \omega_{1/2} \Delta t - \omega_{1/2} e^{-\omega_{1/2} \Delta t} \cos \omega_{1/2} \Delta t + \omega_{1/2}  \\
B &= \omega_{1/2}  e^{-\omega_{1/2} \Delta t} \sin \omega_{1/2} \Delta t + \Delta \omega e^{-\omega_{1/2} \Delta t} \cos \omega_{1/2} \Delta t - \Delta\omega.
\end{aligned}
\end{equation}
The generator current $\boldsymbol{\tilde{I}_g}(t)$ is continuous. Numerically, for a given initial driving current $\boldsymbol{\tilde{I}_{g}(0)}$, Eq.~\ref{eq:3.7.3} can be solved to obtain $\boldsymbol{\tilde{V}_{g}(t)}$ step by step in time domain.  It is referenced as the generator dynamics. Moreover, Eq.~\ref{eq:3.7.3} is still available even when there exists feedback current $\boldsymbol{\delta \tilde{I}_{g}(t)}$.

The beam current $\boldsymbol{\tilde{I}_b}(t)$ is discrete and can be expressed as  
\begin{equation}
\label{eq:3.7.5}
\boldsymbol{\tilde{I}_b}(t) \propto \sum_{k=-\infty}^{k=\infty}\delta(t - k T_{rf}) =I_{DC} + 2 I_{DC}\sum_{k=1}^{k=\infty} \exp(i k \omega t),  
\end{equation}
where $I_{DC}$ is the DC beam current. It is the beam image current playing a role as the driving term in Eq.~\ref{eq:3.7.3}, which induces $\boldsymbol{\tilde{V}_b}(t)$. Noticeably, only the frequency component at $\omega_{rf}$ is synchronized with the bunch repetition rate so that the driving current due to the beam can be simplified to $-2 I_{DC}\exp(i \omega_{rf} t)$  further. In principle, the beam-induced voltage $\boldsymbol{\tilde{V}_{b}(t)}$ can be obtained by solving Eq.~\ref{eq:3.7.2} as well. However, from the fundamental theory of beam loading, this procedure can be significantly simplified. Assume there is a bunch with charge $q$ passing through the cavity at time $t-\Delta t$, then the beam-induced voltage follows 
\begin{equation}
\label{eq:3.7.6}
\boldsymbol{\tilde{V}_{b}}(t) = (\boldsymbol{\tilde{V}_{b}}(t - \Delta t) + \boldsymbol{\tilde{V}_{b0}/2)}]\exp(\alpha \Delta t), \quad
\alpha = -\frac{1}{\tau_{f}}(1-i \tan{\Psi})
\end{equation}
where  $\Psi$ is the cavity de-tuning angle. The phase and amplitude of the beam induced voltage $\boldsymbol{\tilde{V}_{b0}/2}$ is $\pi$ and $|\boldsymbol{\tilde{V}_{b0}}|=q \omega_{r} R_{L} / Q_{L}$. The accumulated beam-induced voltage $\boldsymbol{\tilde{V}_{b}}(t)$ adds an impulse $\boldsymbol{\tilde{V}_{b0}/2}$ whenever a charged bunch passes by, then it decays and rotates by a factor of $\exp(\alpha \Delta t)$ until the next bunch comes. The information on the bunches at previous turns is not needed.

\begin{figure}[!htp]
\centering
\includegraphics[width=1\linewidth]{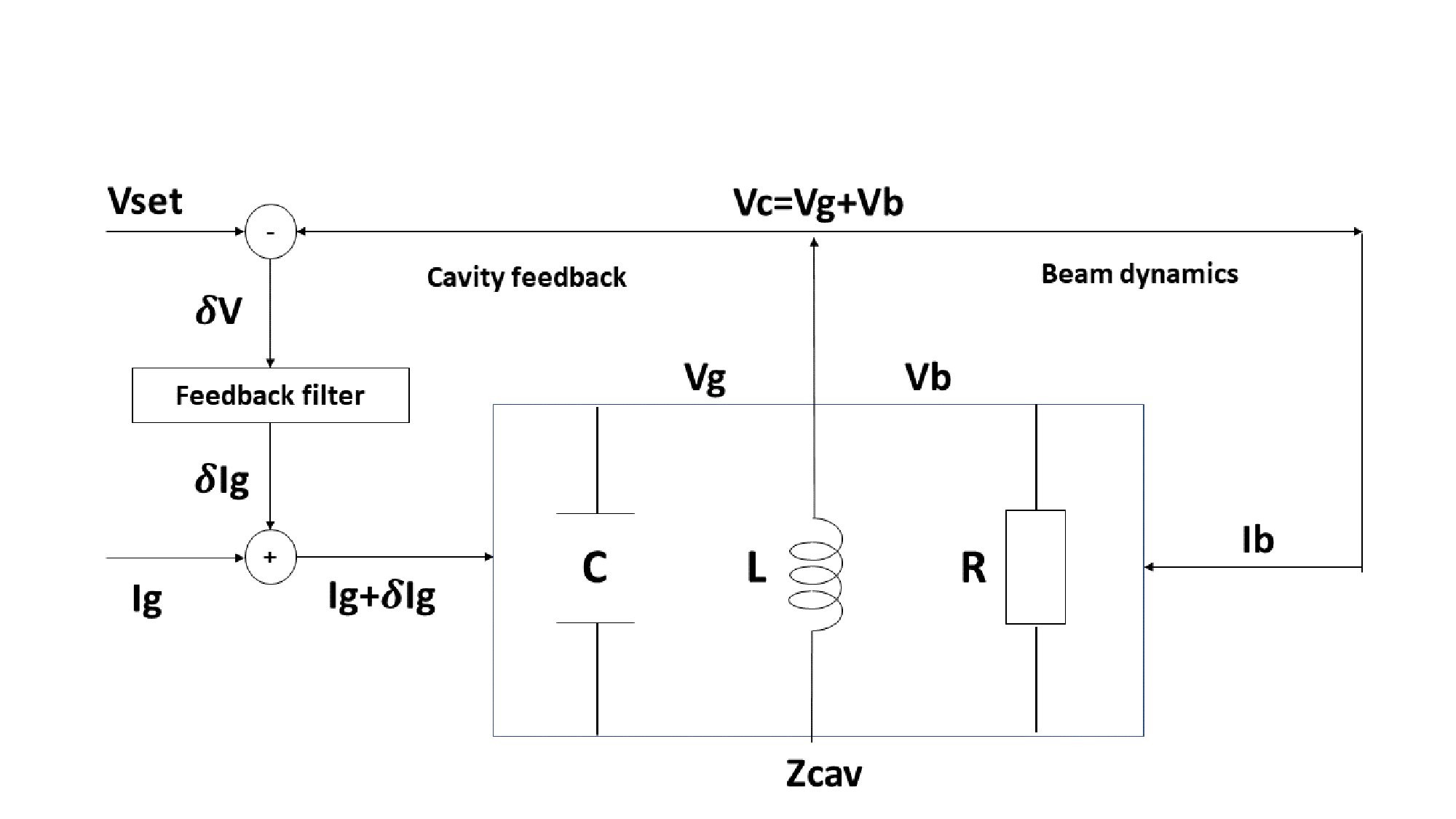}  
\caption{Interaction between generator dynamics, beam dynamics and cavity feedback.}
\label{fig:beam_cavity_feedback}
\end{figure}

In an accelerator with the multi-RFs, the total cavity voltage  \footnote{In the study of transient beam loading effect, the $\cos$ convention is adopted, whereas in Eq.~\ref{eq:3.1.1}, it is $\sin$  convention.}  beam samples usually deviate from the designed value due to the beam-induced voltage, especially when the ring is filled non-uniformly,
\begin{equation}
\label{eq:3.7.7}
\begin{aligned}
\boldsymbol{\tilde{V}_c}(t) = \sum_n (\boldsymbol{\tilde{V}_{g,n}}(t) + \boldsymbol{\tilde{V}_{b,n}}(t)) \neq \sum_n V^{set}_{c,n} \exp(i\omega_{rf,n} t + \phi_n). 
\end{aligned}
\end{equation}
Different bunches will sample different cavity voltages and phases. Thus, cavity feedforward, or feedback, is required to compensate the transient beam-loading effect. Fig.~\ref{fig:beam_cavity_feedback} shows a loop of cavity feedback at the generator side. The performance of the cavity feedback systems is determined by the cavity phasor  $\boldsymbol{\tilde{V}_{c,n}}(t)$ measured at RF cycle $t=m T_{rf}$. The cavity feedback filter could be a $Finite$ $Impulse$ $Response$ (FIR) or a $Infinite$ $Impulse$ $Response$ (IIR) types, which takes the generator current $\boldsymbol{\tilde{I}_{g,n}}(t)$ and 
cavity voltage error $\boldsymbol{\delta \tilde{V}}(t)=\boldsymbol{\tilde{V}^{set}_{c,n}}(t)-\boldsymbol{\tilde{V}_{c,n}}(t)$ as input and gives $\delta \boldsymbol{\tilde{I}_{g}}(t)$ to be modified at the generator side as output. Explicitly, the cavity feedback can be expressed as  
\begin{equation}
\label{eq:3.7.8}
\boldsymbol{\delta\tilde{I}_{g}}(n T_{rf}) =-\frac{1}{a_0}\sum_{i=1}^{N}a_i \boldsymbol{\delta\tilde{I}_{g}}((n-i)T_{rf})+ \frac{1}{b_0}\sum_{j=0}^{M}b_j \delta \boldsymbol{\tilde{V}}((n-j)T_{rf}).
\end{equation}

Finally, we summarize the steps for the transient beam loading simulation in one turn,
\begin{enumerate}
\item Solve Eq.~\ref{eq:3.7.3} to get the $\boldsymbol{\tilde{V}_{g,n}}(t)$; 
\item Solve Eq.~\ref{eq:3.7.6} to get the $\boldsymbol{\tilde{V}_{b,n}}(t)$;
\item Solve Eq.~\ref{eq:3.7.7} to get the $\boldsymbol{\tilde{V}_c,n}(t)$;
\item Transfer the particle by the one-turn map;
\item Solve Eq.~\ref{eq:3.7.8} to get the $\boldsymbol{\delta \tilde{I}_{g,n}}(t)$.
\end{enumerate}

One last thing to emphasize, the above discussion assumes a zero bunch length. This approximation is not appropriate for a bunch with a finite length. In that case, the bunch can be cut into different bins in the time sequence, thereafter, the bins of each bunch can be treated as zero-length micro-bunches, which pass through the cavities one after another. Then the above equations are still available.

\section{4 CETASim benchmarks and implementation}
\begin{figure}[!htp]
\includegraphics[width=1\linewidth]{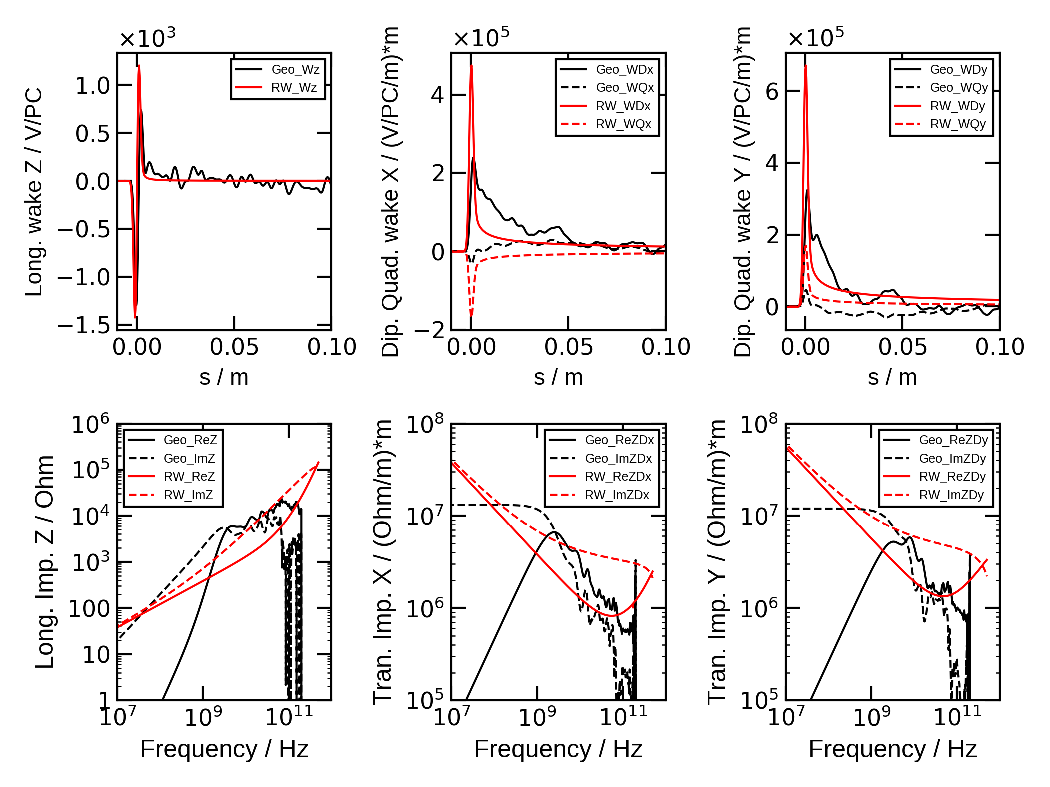}
\caption{Above: wakes of a 1 $mm$ leading electron Gaussian bunch due to the geometrical impedance and the resistive wall impedance. Bottom: the geometrical impedance and the resistive wall impedance. From left to right, figures correspond to the longitudinal, horizontal and vertical directions respectively. The quadruple impedance is not plotted here.}
\label{fig:4.1}
\end{figure}

The H6BA lattice of the PETRA-IV storage ring \cite{Petra4CDR} is applied to show the capability of CETASim. PETRA-IV will adopt an active 3rd harmonic cavity to lengthen the bunch. The ideal bunch lengthening condition requires $Re(\boldsymbol{\tilde{V}_c}(\tau=0))=U_0/e$ 
to compensate the radiation loss $U_0$, $Re(\boldsymbol{\tilde{V}_c'}(\tau=0))=Re(\boldsymbol{\tilde{V}_c''}(\tau=0))=0$ to have a flat RF potential. The broadband impedance comprises the RW impedance and the geometric impedance. The RW impedance is created by the resistive chambers of the IDs, as well as the rest of the ring. The geometrical impedance is due to the elements in the ring such as BPMs, Bellows, Flanges, $etc$. The program ImpedanceWake2D \cite{bib:IW2D} is used to compute the resistive wall impedance. GdfidL computes the geometric wake potentials \cite{bruns2002gdfidl}, where a 1 $mm$ long Gaussian bunch is used as the driving bunch. Fig.~\ref{fig:4.1} shows the wake potential in the longitudinal, horizontal, and vertical planes of a 1 $mm$ leading electron Gaussian bunch and the impedance correspondingly. Here, the transverse wake potential and impedance are multiplied by the average betatron function $\langle\beta\rangle$ so that the unit differs from conventional ones.

\begin{table}[!htp]
\caption{Nominal lattice parameters of PETRA-IV H6BA lattice.}
\centering
\begin{tabular}{|l|r|r|r|r|}  
\hline
Parameters   &units    &symbol       &  DW Closed  & DW Open \\
\hline
Energy      &GeV       &$E$          & 6           & 6     \\
\hline
Circumference  &m                &$C$          & 2304        & 2304  \\
\hline
Natural Emittance  &pm          &$\epsilon_0$   & 20          & 43    \\
\hline
Emittance Ratio &          &$\kappa$         & 0.1         & 0.1   \\
\hline
Tunes           &          &$\nu_x$/$\nu_y$  & 135.18/86.27  & 135.18/86.23   \\
\hline
Momentum Compact Factor &         &$\alpha_c$            & 3.33$\times 10^{-5}$    & 3.33$\times 10^{-5}$   \\
\hline
Damping Time  &ms        &$\tau_x$ & 17.76      &  39.23   \\
\hline
Damping Time  &ms        &$\tau_y$ & 22.14      & 69.63    \\
\hline
Damping Time  &ms        &$\tau_s$  & 12.62      & 56.84    \\
\hline
Natural Energy Spread &rad        &$\sigma_E$    &  8.9$\times 10^{-4}$    & 7.37$\times 10^{-4}$  \\
\hline
Natural Bunch Length  &mm     &$\sigma_s$ & 2.3        & 1.794  \\
\hline
Energy Loss    &MeV        & $U_0$  & 4.166      & 1.423  \\
\hline
Main Cavity Voltage &MV        &$V_{c,1}$ & 8          & 8      \\
\hline
Main Cavity Harmonics &      &$h$        & 3840       & 3840    \\
\hline
\end{tabular}
\label{tab:4.1}
\end{table} 

\subsection{4.1 Zero current beam equilibrium state}
\begin{figure}[!htp]
\centering
\includegraphics[width=1\linewidth]{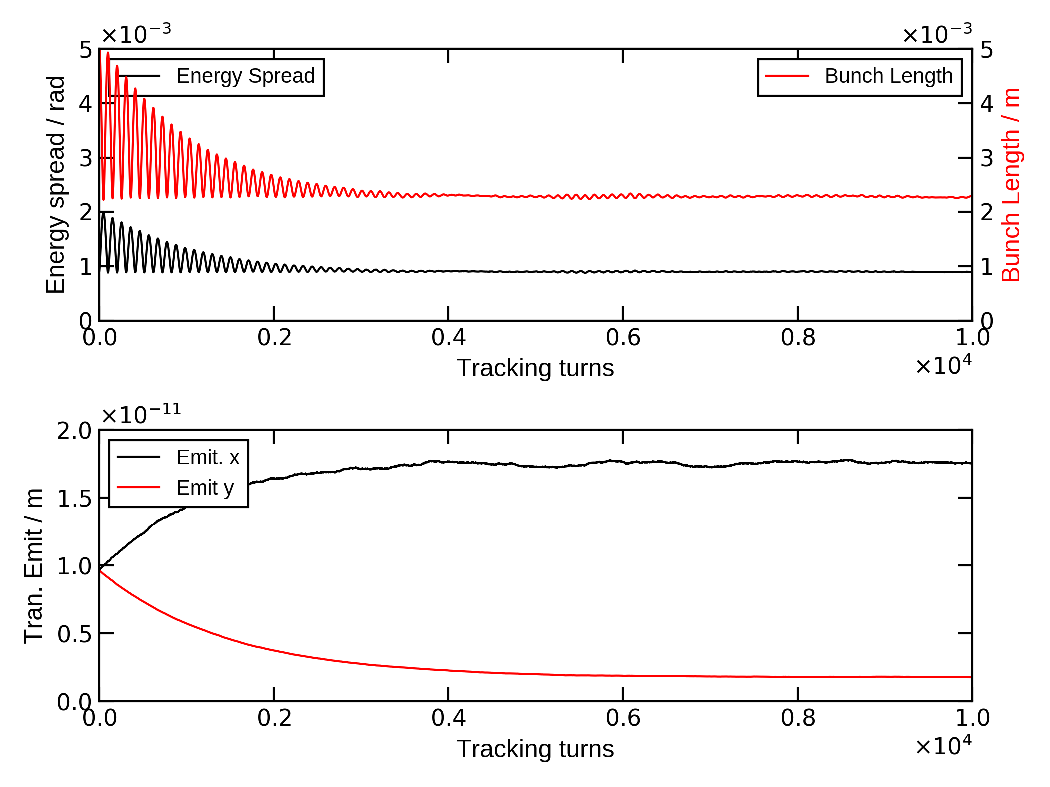} 
\caption{Bunch energy spread, bunch length (top) and transverse beam emittance (bottom) as a function of tracking turns. The beam evolves to the natural equilibrium state as expected.}
\label{fig:4.0}
\end{figure}

Firstly, we would like to show the influence of radiation damping and quantum excitation. We prepare one bunch with an initial Gaussian distribution both in the longitudinal and transverse phase space. $10^4$ macro-particles are generated. The initial transverse bunch emittance is set as $\epsilon_{x,0}=\epsilon_{y,0}=10$ $pm$, and the initial energy spread and bunch length are set as $\sigma _{E,0}=8.9\times 10^{-4}$ $rad$, $\sigma _{z,0}=5.3\times 10^{-3}$ $m$. The emittance ratio $\kappa$ in the transverse plane and the radiation damping times shown in Tab.~\ref{tab:4.1} are adopted in simulation. Fig.~\ref{fig:4.0} shows how the bunch length, energy spread and the transverse emittance evolve as a function of tracking turns. The beam evolves to the natural equilibrium state as expected.

\subsection{4.2 Single bunch effect}
\begin{figure}[!htp]
\centering
\includegraphics[width=1\linewidth]{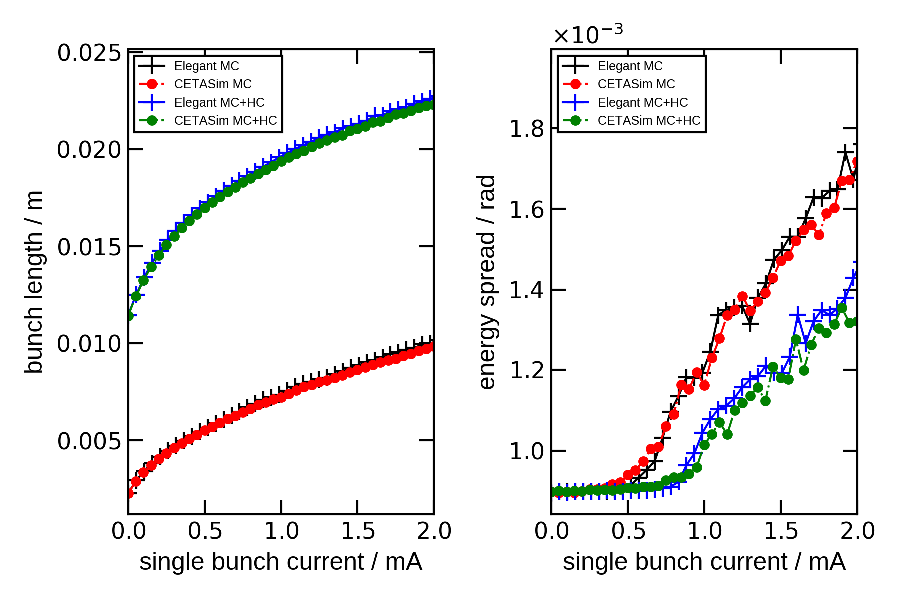} 
\caption{Single bunch length (left), energy spread (right) as a function of bunch current.}
\label{fig:4.2}
\end{figure}

With the broadband impedance shown in Fig.~\ref{fig:4.1}, we perform a study of the single-bunch effects \cite{Petra4TDR}. The longitudinal impedance leads to potential well distortion and bunch lengthening effects. Once the single bunch current increases further and the longitudinal microwave instability threshold is reached that the energy spread starts to increase as well. We give two scenarios studied by CETASim: the main cavity only and the main cavity together with a 3rd harmonic cavity.  In the double RF case, the generator voltage and phase are set to maintain the ideal bunch lengthening condition. Fig.~\ref{fig:4.2} shows the single bunch length and energy spread as a function of the single bunch current. In each sub-figure, two groups of curves are given. The red ones are from CETASim and the black ones are from Elegant. The bunch length given by CETASim agrees well with the results given by Elegant. Noticeably, there are discrepancies in the energy spread when the single bunch current is above the longitudinal microwave instability.

\begin{figure}[htp]
\includegraphics[width=1\linewidth]{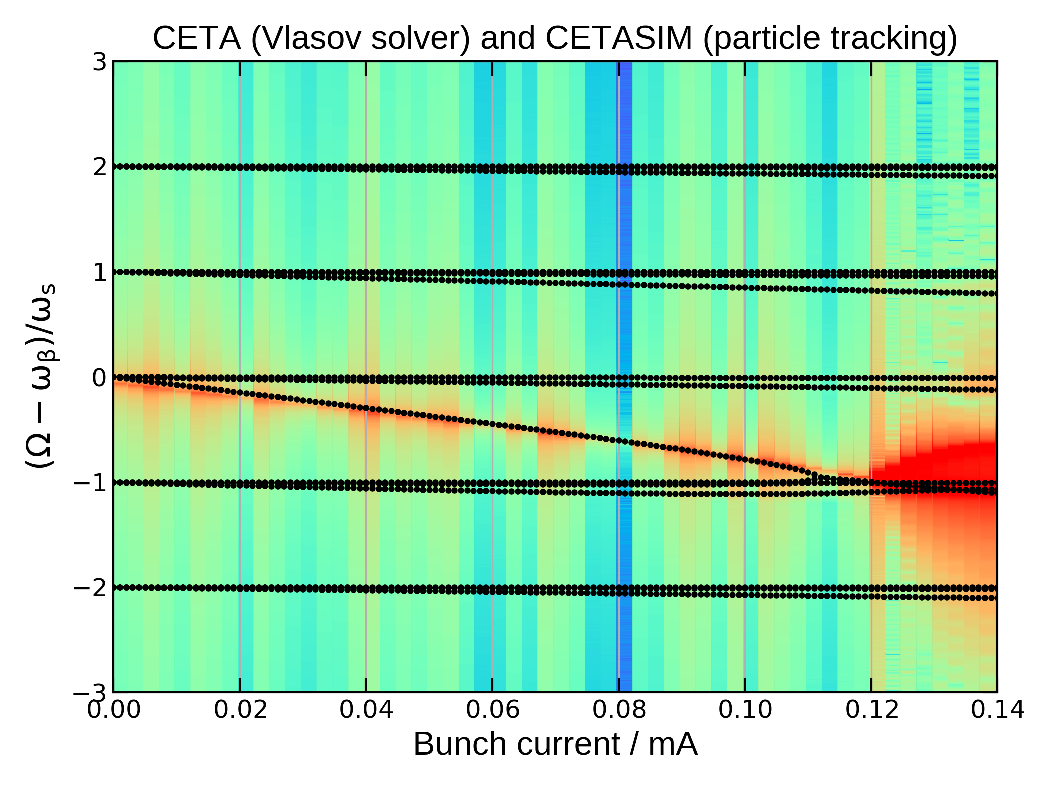} 
\caption{Comparison of the TMCI threshold from Vlasov solver (black dots) and CETASim (contours). In CETASim tracking, only the vertical dipole impedance is taken into account.}
\label{fig:4.3}
\end{figure}

In the transverse plane, the impedance leads to the Transverse Mode Coupling Instability (TMCI) and head-tail instabilities. The azimuthal modes shift as the single bunch current increases. At a certain point, these azimuthal modes collide and merge, then the beam becomes unstable. If there exists dipole impedance only, the mode coupling condition can be found by Sacherer’s method \cite{AlexChao}. Considering the dipole impedance effect as a perturbation and decomposing the longitudinal phase space by the Laguerre polynomials, a set of Sacherer integral equations can be established, by solving which the beam threshold due to mode merge conditions can be found analytically. As a benchmark study, we set the simulation with zero chromaticity and take into account the vertical dipole impedance only. The comparison of the azimuthal mode frequency shift between CETASim tracking and the Vlasov solver is shown in Fig.~\ref{fig:4.3}. The black dots and contours represent results from the Vlasov solver and CETASim respectively, which give consistent TMCI beam current threshold.    

\begin{figure}[!htp]
\includegraphics[width=1\linewidth]{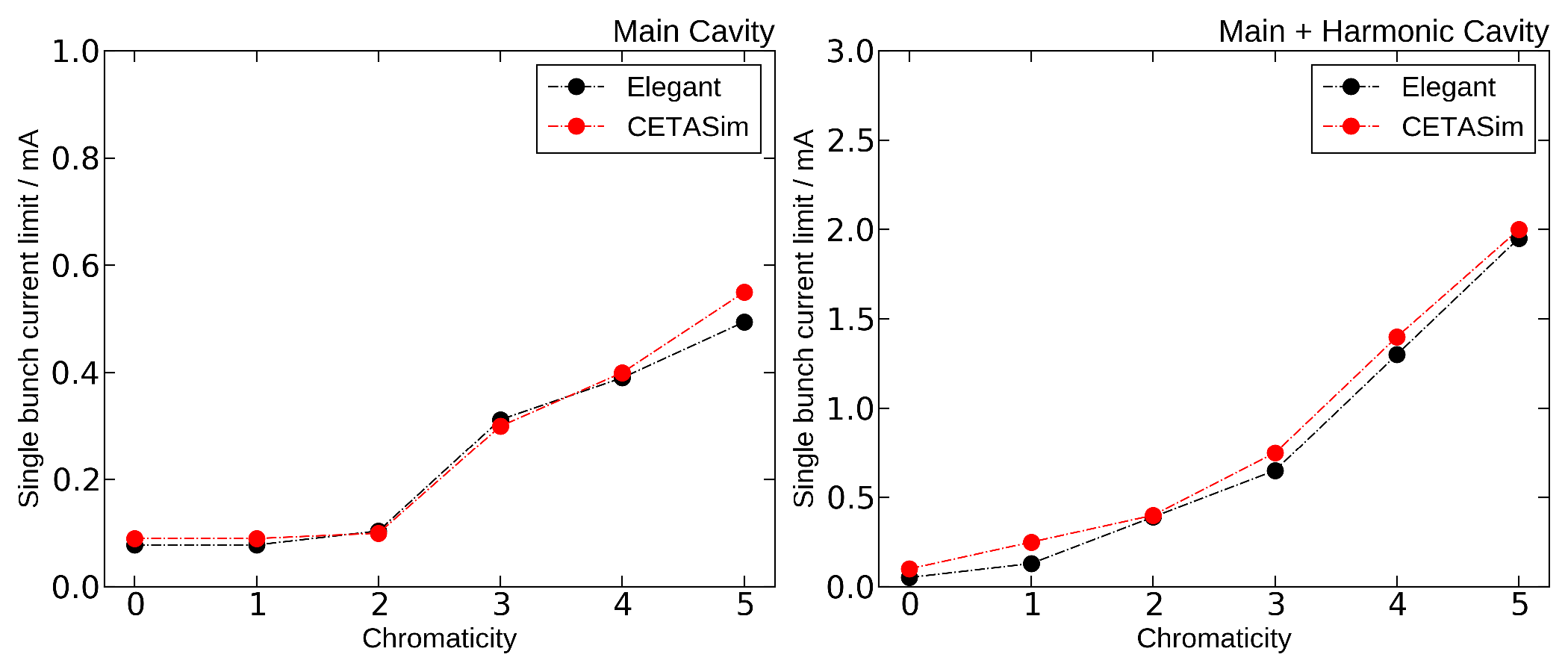} 
\caption{The single bunch current threshold as a function of chromaticity without (left) and with (right) the 3rd harmonic cavity.}
\label{fig:4.4}
\end{figure}

One way to increase the transverse single-bunch current limit is to set a non-zero chromaticity $\xi$. The chromaticity introduces a head-tail phase advance and shifts the mode spectrum by $\omega_{\xi}=\xi f_0/\eta$, where $\eta$ is the lattice slip factor and $f_0$ is the nominal revolution frequency.  Fig.~\ref{fig:4.4} shows the single bunch current limit as a function of chromaticity without and with 3rd harmonic cavity. The transverse dipole, quadrupole impedance, and longitudinal impedance are all taken into account in simulation. The threshold current is defined as the lowest bunch current, which leads to a non-100\% transmission during the tracking, where an elliptical chamber with a radius (15, 10) $mm$ is applied as the particle loss criteria.

\subsection{4.3 Coupled-bunch instability due to the long-range wakes}
In general, if the impedance is known and the filling pattern of the ring is uniform, the coupled bunch mode frequency shift and growth rate can be obtained
\begin{equation}\label{eq:4.2.1}
\begin{split}
(\Omega^{\mu} - \omega_{\beta})_{\perp} &= -i \frac{M N r_0 c}{2 \gamma T_0^2 \omega_{\beta}} \sum_{p=-\infty}^{\infty} Z_1^{\perp}[\omega_{\beta} + (pM + \mu) \omega_0 ] \\
(\Omega^{\mu} - \omega_{s})_{\parallel} &=  i \frac{M N r_0 \eta}{2 \gamma T_0^2 \omega_{s}} \sum_{p=-\infty}^{\infty} (\omega_{s} + pM\omega_0  + \mu \omega_0 ) Z_0^{\parallel}[\omega_{s} +  pM\omega_0  + \mu \omega_0  ], 
\end{split}
\end{equation}
where $M$ is the number of equal spaced bunch number, $N$ is the electron particle number per bunch, $\mu$ is the coupled bunch mode index varying from 0 to $M$-1, $\omega_{\beta}$ and  $\omega_{s}$ are the transverse and longitudinal oscillation angular frequency. In particle tracking studies, the coupled bunch mode can be reconstructed when the bunch-by-bunch and turn-by-turn data is available. Take the horizontal plane for example, the coupled bunch mode can be obtained from the Fourier spectrum of the one-turn complex signal $z_\mu$ at a certain BPM
\begin{equation}\label{eq:4.2.2}
z_{\mu} = (\frac{x_\mu}{\sqrt{\beta_x}} - i (\sqrt{\beta_x} p_{x,\mu} + \frac{\alpha_x}{\sqrt{\beta_x}} x_\mu ) e^{-i\frac{2\pi \nu_x(\mu-1)}{M}},
\end{equation}
where $\beta_x$ and $\alpha_x$ are the Twiss parameters at the BPM. The growth rate of the $\mu$th mode can be obtained by an exponential fitting of the $\mu$th mode spectrum amplitude as a function of tracking turns.

However, in a real machine, what can be measured at the BPMs is limited to the position information only. In CETASim, we also supply another approach to get the coupled bunch instability growth rate by following the process of drive-damp experiments \cite{rehm2014new}. A driver can be set in CETASim to excite the beam with a given kick strength and frequency $f_{\mu}$ by which the $\mu$th coupled bunch mode can be excited. Then the signal
\begin{equation}\label{eq:4.2.3}
\tilde{S}^{\mu}_{n} = \sum_{m=0}^M \tilde{s}^{\mu}_{n,m} = \sum_{m=0}^M  x_{n,m} \exp(- i 2\pi  f_\mu h T_{rf} / M ),
\end{equation}
is recorded turn by turn. The growth rate of the excited $\mu$th mode can be found by an exponent fitting of $|\tilde{S}^{\mu}_{n}|$. Then, one can reconstruct the full picture of the coupled-bunch mode by scanning the driving frequency $f_{\mu}$ of the driver.

\begin{figure}[!htp]
\includegraphics[width=1\linewidth]{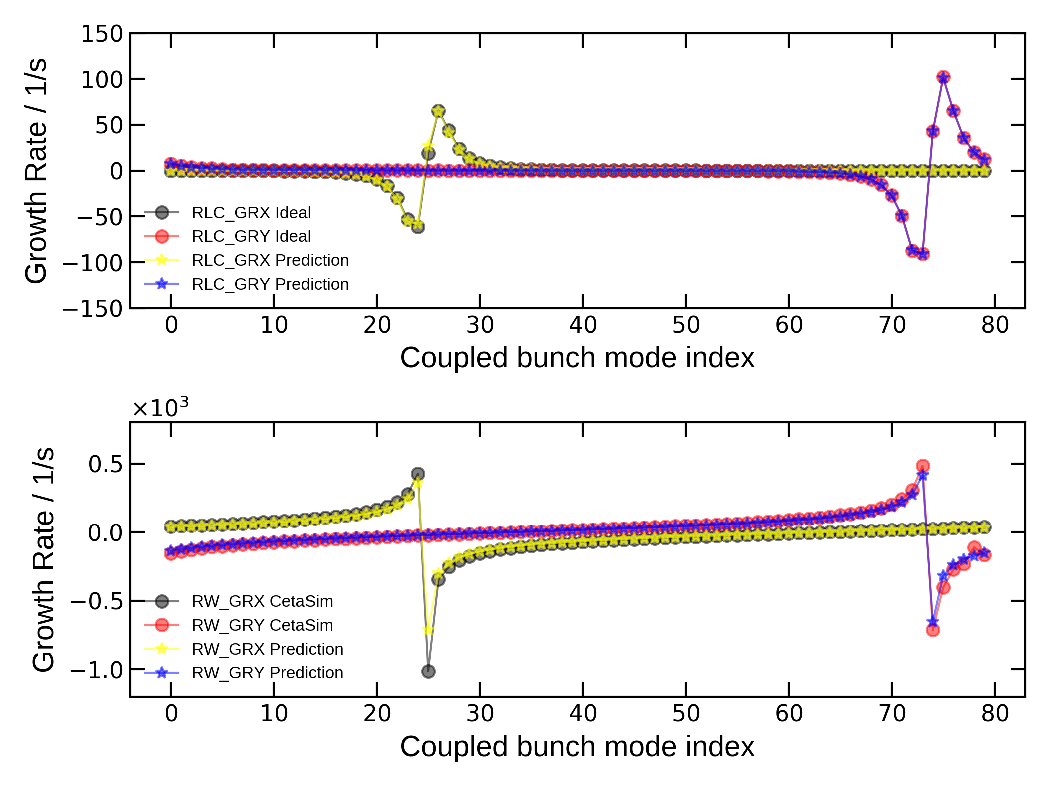}
\caption{Growth rate of the transverse coupled bunch modes from RLC (top) and RW (bottom) wakes. The RLC parameters are $R_s=5.\times 10^9$ $ohm/m^2$, $Q=1.\times 10^3$ and $\omega_r=2\pi \times 4.996 \times 10^8$ $1/s$. The RW parameters are shown in Tab.~\ref{tab:4.2}. The ring is filled uniformly by 80 bunches and each bunch current is 1 $mA$. The long-range wakes last 20 turns and the SR damping is turned off. 'Ideal' and 'Prediction' indicate the results are obtained from Eq.~\ref{eq:4.2.2} and Eq.~\ref{eq:4.2.1} respectively.}
\label{fig:4.5}
\end{figure}

In CETASim, there are two analytical models to generate the long-rang wakes, the RLC model and the RW model. In below, two examples are given for the transverse coupled bunch effect study. In the first case, the transverse impedance modeled by an RLC resonator with the parameters $R_s=5 \times 10^9$ $\Omega/m^2$, $Q=1\times 10^{-3}$ and $\omega_r=2\pi \times 4.996 \times 10^8$ $1/s$. The second one is a simplified resistive wall impedance of the PETRA-IV storage ring, where the vacuum chamber is grouped into 4 types of sections as shown in Tab.~\ref{tab:4.2}. The characteristic distance $z_0$ of these four sections are in the order of $1.\times 10^{-5}$ m, which is much smaller than the RF bucket distance $c T_{rf}=0.6$ m, so that Eq.~\ref{eq:3.2.5} applied in CETASim is still a good approximation. In the simulation, the ring is filled uniformly by 80 bunches and each bunch has 1 $mA$ current. The long-range wakes are truncated at the 20$th$ turn. The synchrotron radiation damping is turned off. In Fig.~\ref{fig:4.5}, we give the results of the coupled bunch mode growth rate from CETASim tracking ('Ideal') and analytical predictions ('Prediction'). The results obtained from tracking show very good agreements with predictions for both RLC and RW impedance.

\begin{table}[!htp]   
\caption{Simplified resistive wall sections in PETRA-IV.}
\centering
\begin{tabular}{|l|r|r|r|r|r|r|r|} 
\hline
section  &number  &len. / m    & gaps / mm & $\beta_x$ / m & $\beta_x$ / m & conduct. $\sigma$ / $\Omega^{-1} m^{-1}$ & $z_0$ / m \\
\hline
5 mm ID      &  4      &  5        & 5      & 3.14  & 3.14  &  2.5$\times 10^7$ & 1.74$\times 10^{-5}$ \\
\hline
6 mm ID      &  17     &  5        & 6      & 3.14  & 3.14  &  2.5$\times 10^7$ & 1.97$\times 10^{-5}$ \\
\hline
7 mm ID      &  5      &  10        & 7      & 6.08  & 6.08  & 2.5$\times 10^7$ & 2.18$\times 10^{-5}$ \\
\hline
ring         &  1      &  2149       & 10     & 2.71 & 4.25  & 5.9$\times 10^7$ & 2.08$\times 10^{-5}$\\
\hline
\end{tabular}
\label{tab:4.2}
\end{table}

\begin{figure}[!htp]
\includegraphics[width=1\linewidth]{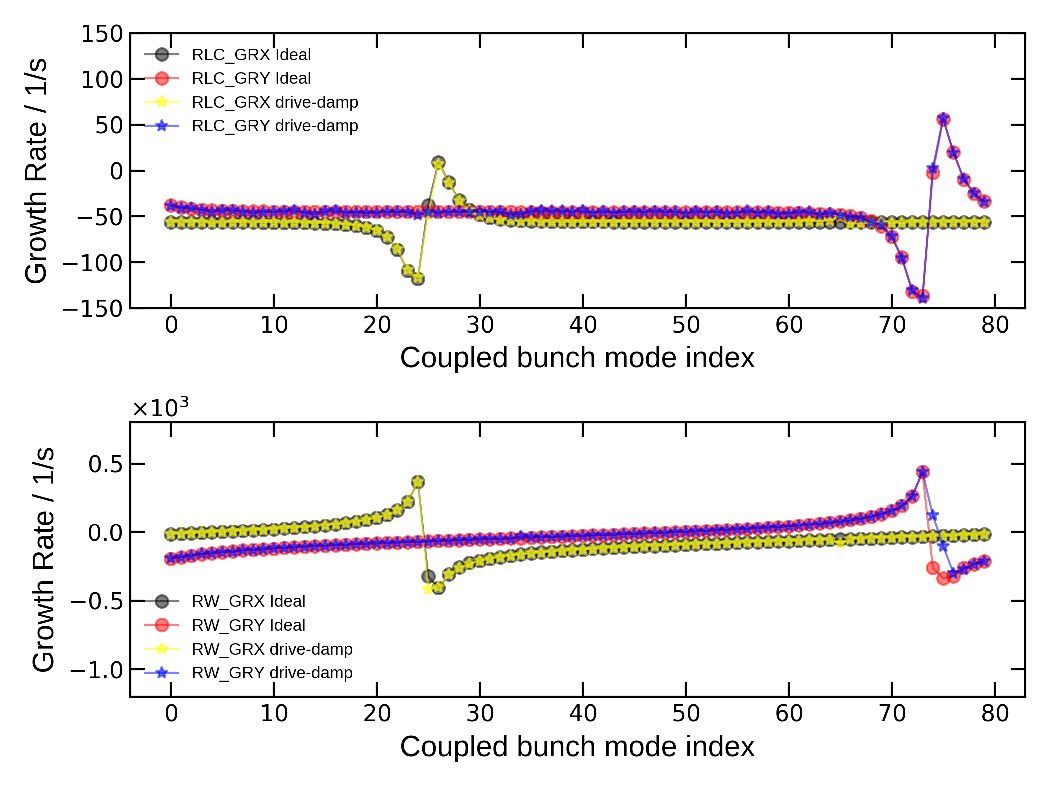} 
\caption{Growth rate of the transverse coupled bunch modes from RLC (top) and RW (bottom) wakes. The long-range wakes last 20 turns and the synchrotron radiation damping is turned on. 'Drive-damp' indicates the results are obtained from Eq.~\ref{eq:4.2.3}.}
\label{fig:4.6}
\end{figure}

Figure~\ref{fig:4.6} represents the coupled bunch modes reconstructed from the drive-damp method. The driving strength of the exciter is 0.2 $\mu$rad and the excitation time is limited to the first 300 turns. The bunch-by-bunch data of the following 700 turns are used for coupled bunch mode growth rate calculation. The simulation conditions are the same as those used in Fig.~\ref{fig:4.5}, except the radiation damping is turned on. For a better comparison, the results obtained from Eq.~\ref{eq:4.2.2} (‘Ideal’) are plotted as well. Both methods give the same coupled bunch growth rate. Compared to results in Fig.~\ref{fig:4.5}, the growth rates are decreased by turning on the radiation damping. 

In the longitudinal plane, the tracking studies with the longitudinal long-range wakes can be set up similarly in CETASim. 

\subsection{4.4 Transient beam loading effect}
\begin{table}[htp]   
\caption{RF parameters of PETRA-IV storage ring.}
\centering
\begin{tabular}{|l|r|r|r|} 
\hline
Parameter  &Symbol  &Main RF ($n=1$)    &Harmonic RF ($n=3$) \\
\hline
RF Freq. (Hz)     & $f_{rf,n}$       & 4.996$\times 10^8$    & 1.499$\times 10^9$     \\ 
\hline
Ref. Voltage (V)  & $V_{c,n}$        & 8$\times 10^6$        & 2.223$\times 10^6$     \\
\hline
Synchronous Phase (Rad)     & $\phi_{n}$       & 2.516       & -0.236       \\
\hline
Detuning Freq. (Hz)    & $\Delta f_{n}$   & -27.9$\times 10^3$    & 277.6$\times 10^4$     \\ \hline
Coupling Factor         & $\beta_n$        & 3.0        & 5.3       \\
\hline
Shunt Impedance         & $R_{s,n}$        & 81.6$\times 10^6$    & 36.$\times 10^6$    \\
\hline
Quality  Factor         & $Q_{0,n}$        & 29600      & 17000     \\
\hline
Gen. Curr. Amp. (Amp.)   & $Abs$($\boldsymbol{\tilde{I}_g}$)  & 0.626  & 0.294 \\
\hline
Gen. Curr. Phase (Rad)   & $Arg$($\boldsymbol{\tilde{I}_g}$)  & 0.945  & -1.813 \\
\hline
Total Generator Power (W)   & $P_{g,n}$     & 1.34$\times 10^6$    & 7.40$\times 10^4$    \\
\hline
Total Reflected Power (W)   & $P_{r,n}$     & 4.98$\times 10^3$    & 1.09$\times 10^5$    \\
\hline
\end{tabular}
\label{tab:4.3}
\end{table}

\begin{figure}[htp]
\includegraphics[width=1\linewidth]{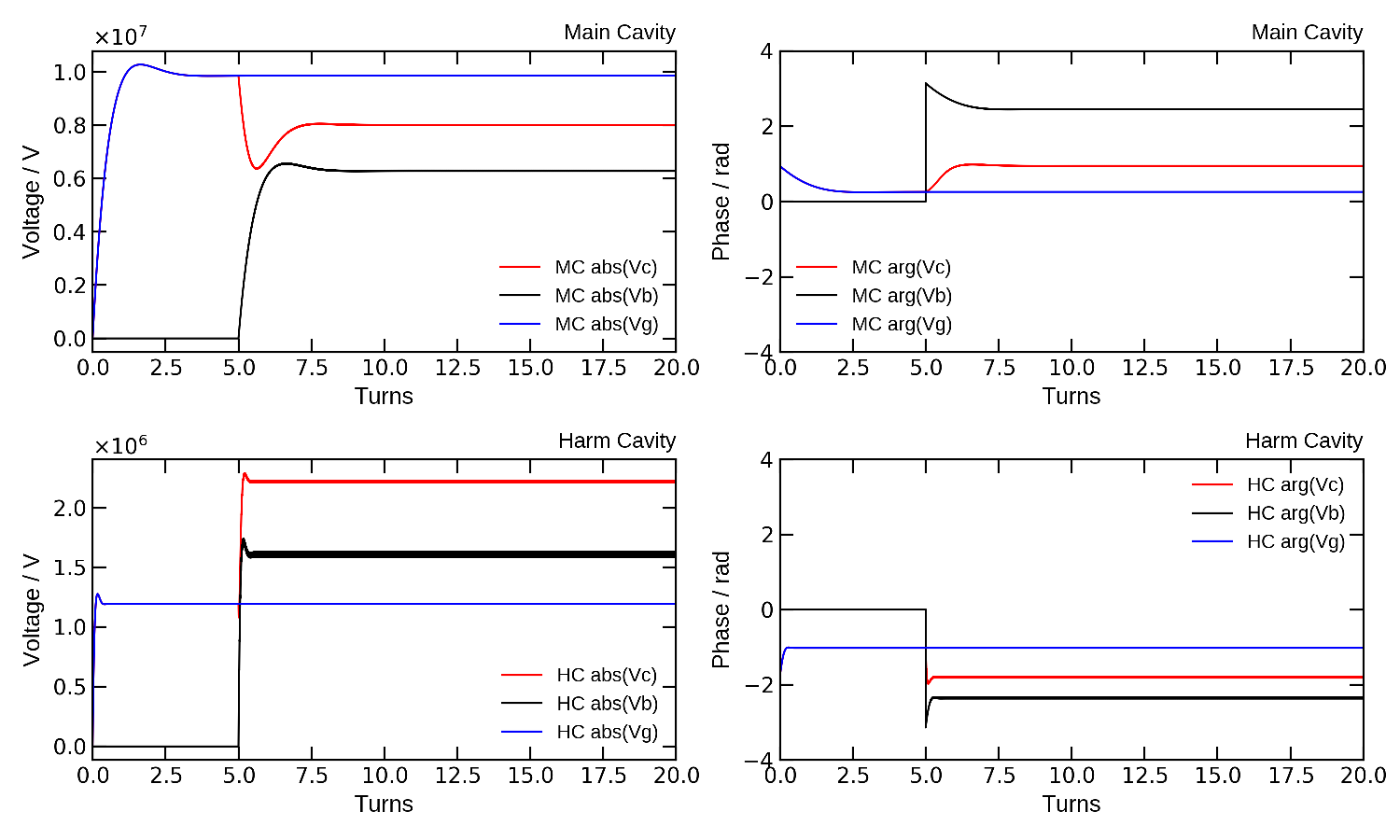} 
\caption{The amplitude (left) and phase (right) of beam-induced voltage $\boldsymbol{\tilde{V}_b}$, generator voltage $\boldsymbol{\tilde{V}_g}$ and cavity voltage $\boldsymbol{\tilde{V}_c}$ on the main cavity (top) and harmonic cavity (bottom) as function of turns. The beam current is taken into account from the 5$th$ turn.}
\label{fig:4.7}
\end{figure}

Transient beam loading brings two effects: the longitudinal coupled bunch instability and an unexpected bunch lengthening. In CETASim, the coupled bunch instability can be turned off by ignoring the dynamical variation of the beam-induced voltage. In that case, the tracking results always converge to an equilibrium state. If tracking is set up as that one bunch is composed of one macro-particle, according to the phase and voltage  bunches sampled, the bunch profiles can be found analytically
\begin{equation}
\label{eq:4.3.1}
\begin{split}
\rho(z) &= \rho_0 \exp(-\frac{1}{2\pi h f_0 \alpha_c \delta^2}H_1(z)), \\
H_1(z)  &= \frac{\omega_0 e}{2 \pi \beta^2 E} \frac{2 \pi h f_0}{\beta c} (Re \sum_n \int_0^{z}  \boldsymbol{\tilde{V}_{c,n}}(z')dz' + \int_0^{z} \int_{z''}^{\infty} e\rho(z') W_0'(z''-z')dz'dz''). 
\end{split}
\end{equation}
Here, $H_1(z)$ is the Hamiltonian composed of terms due to the RF potential and the short-range wakes \cite{AlexChao}. If multi-particles are set per bunch, then the bunch profile can be obtained by binning the distribution longitudinally in real space.  Tab.~\ref{tab:4.3} gives the RF parameters of the main cavity and the 3rd harmonic systems in PETRA-IV. The cavities are de-tuned to the 'optimized' condition to have the lowest power consumption. The filling pattern of the ring is set as $h=3840=80\times(20\times(1+1)+8)$. There exist 80 bunch trains and in each bunch train, every bucket is occupied by the electron beam except the last 8 buckets. The total beam current is 200 $mA$. Here, we give the simulation results by turning off the longitudinal coupled bunch instability. Fig.~\ref{fig:4.7} shows how the amplitude and phase of the beam induced voltage $\boldsymbol{\tilde{V}_b}$, generator voltage $\boldsymbol{\tilde{V}_g}$ and cavity voltage $\boldsymbol{\tilde{V}_c}$ are built up as functions of turns. With the optimum de-tunning condition, the generator voltage driven by $\boldsymbol{\tilde{I}_g}$ is simulated from the 0$th$ turn, whereas the beam-induced voltage driven by $\boldsymbol{\tilde{I}_b}$ is simulated from the 5$th$ turn. The cavity voltage $\boldsymbol{\tilde{V}_c}$ beam supposed to sample converges to the designed values as expected. 

\begin{figure}[htp]
\includegraphics[width=1\linewidth]{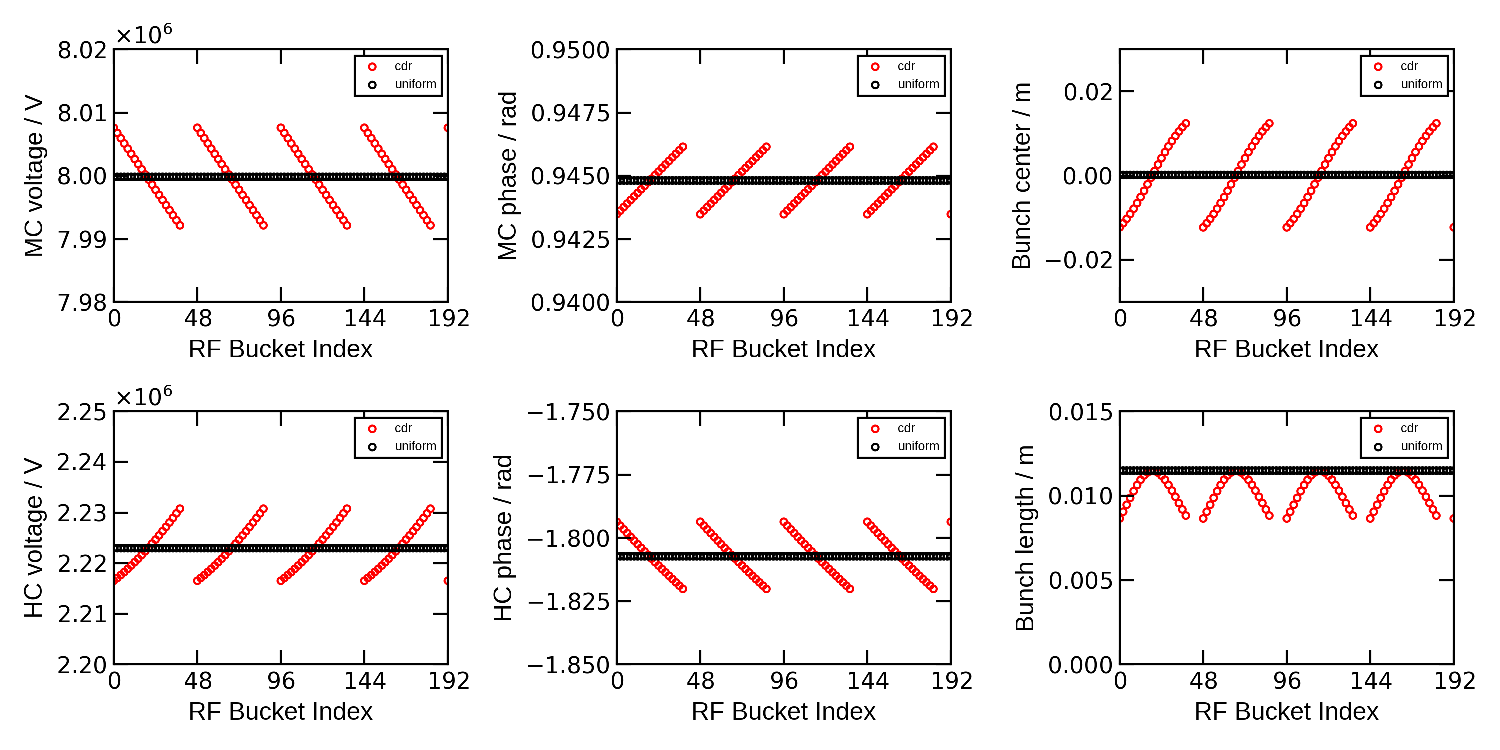} 
\caption{The cavity voltage $Abs(\boldsymbol{\tilde{V}_c})$ (left), cavity phase $Arg(\boldsymbol{\tilde{V}_c})$ (middle) bunch sampled as functions of RF bucket index at the main cavity (top) and harmonic cavity (below); the right column gives the bunch center and bunch length as functions of RF bucket index. The total beam current is 200 $mA$.}
\label{fig:4.8}
\end{figure}

\begin{figure}[htp]
\includegraphics[width=1\linewidth]{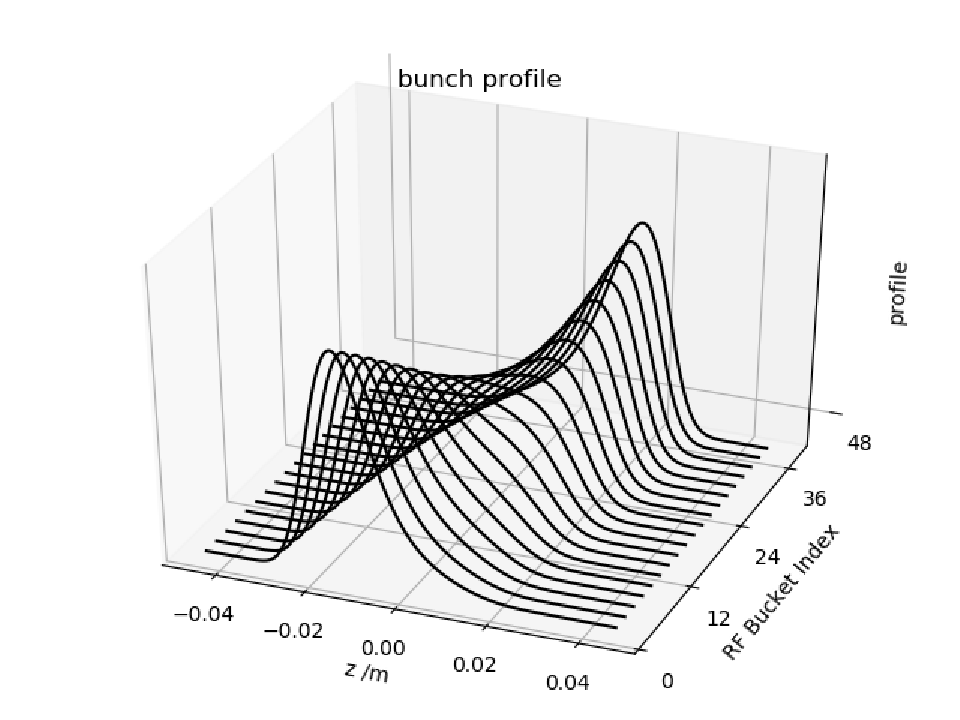} 
\caption{Bunch profiles in the first bunch train as a function of RF bucket index. The profile are obtained from Eq.~\ref{eq:4.3.1}.}
\label{fig:4.9}
\end{figure}

Figure~\ref{fig:4.8} shows the simulation results by setting one macro-particle per bunch. The sub-figures on the left and middle depict the cavity voltage and phase sampled by the bunches in the first 4 bunch trains. The sub-figures on the right show the bunch center shift and bunch length variation. The periodic filling pattern leads to a periodical voltage sampling, which further reduces to a periodical bunch center offset and bunch lengthening effect. The bunch profile obtained from Eq.~\ref{eq:4.3.1} as a function of the RF bucket index is shown in Fig.~\ref{fig:4.9}.  The result suggests that the bunch lengthening is very sensitive to the phase variation in cavities.

We would like to introduce a simple equation \cite{boussard1995beam}, which can be used to estimate the cavity phase modulation $\delta \phi_{max}$ due to the empty gaps $\delta t$ in a bunch train, 
\begin{equation}
\label{eq:4.3.2}
\delta \phi_{max,n} = \frac{1}{2}\frac{R_{s,n}}{Q_{0,n}}\frac{2\pi f_{rf,n}}{V_{c,n}} I_{DC} \delta t. 
\end{equation}
With the filling pattern $h=3840=80\times(20\times(1+1)+8)$, Eq.~\ref{eq:4.3.2} gives the peak-to-peak phase variation around $0.0037$ and $0.029$ $rad$ for the main cavity and harmonic cavity, which are rather good estimation compared to the results from CETASim simulation shown in Fig.~\ref{fig:4.8}.  

\begin{figure}[htp]
\includegraphics[width=1\linewidth]{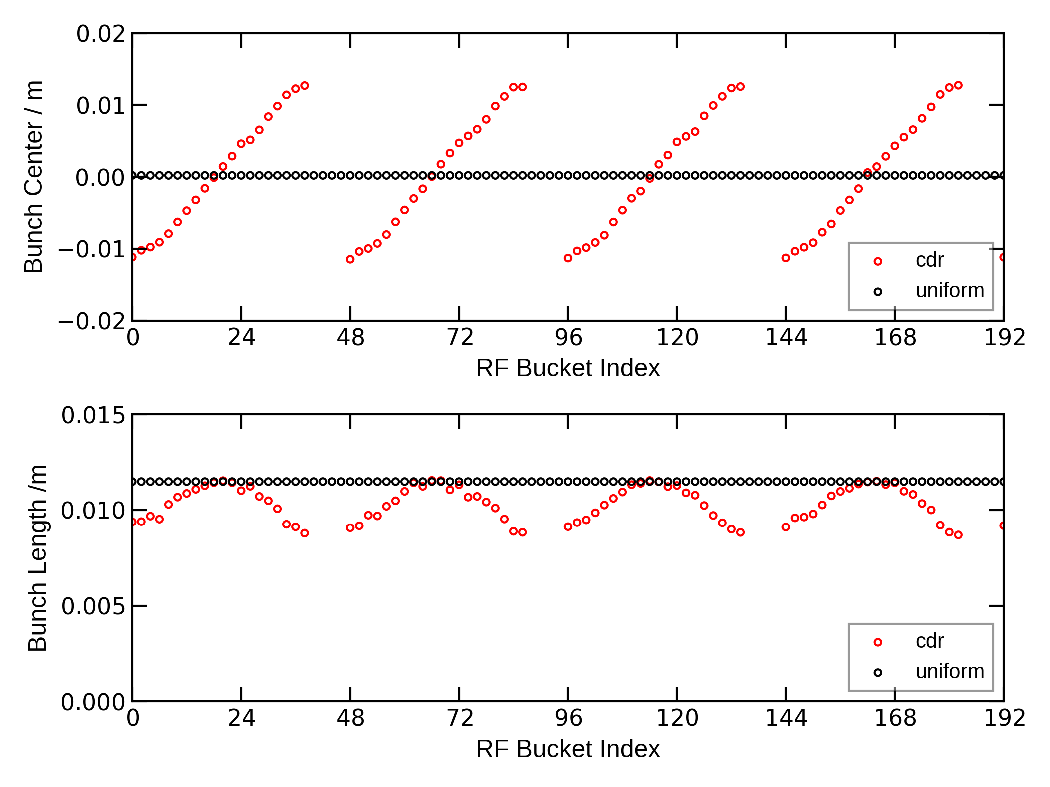} 
\caption{Bunch center and bunch length variation of the first 4 bunch trains as a function of RF bucket index. The results are obtained from multi-particle tracking and each bunch is composed of 3000 macro-particles.}
\label{fig:4.10}
\end{figure}

\begin{figure}[htp]
\includegraphics[width=1\linewidth]{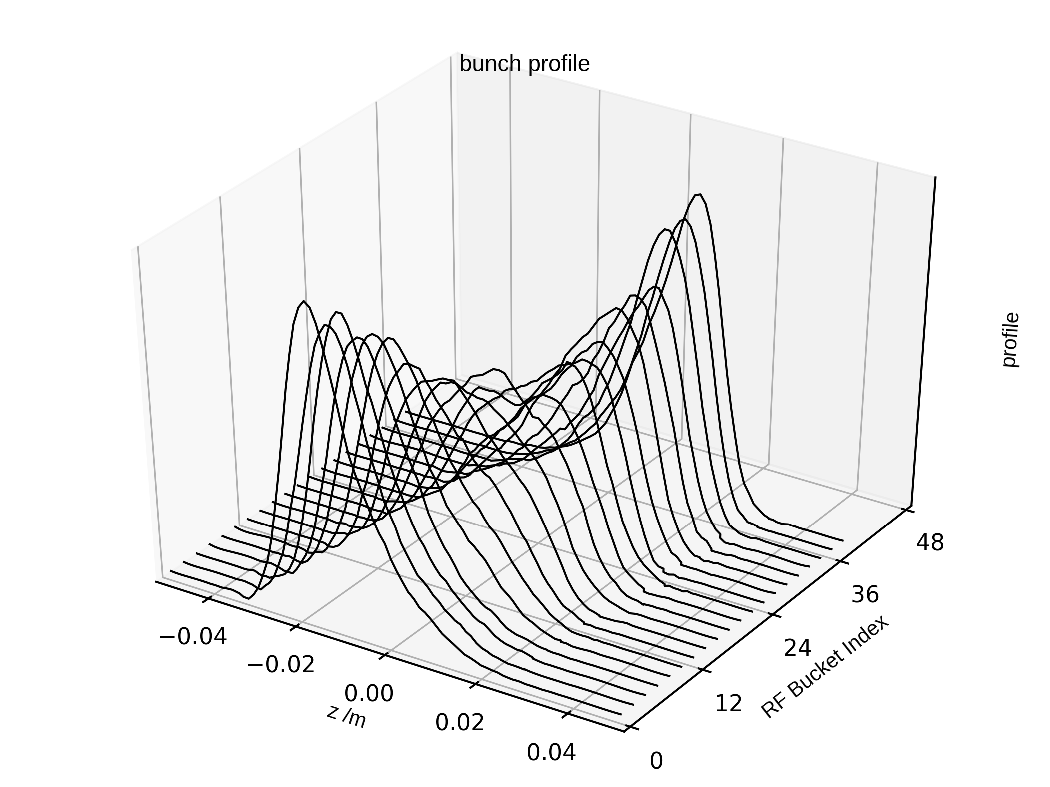} 
\caption{Bunch profiles as a function of RF bucket index in the first bunch train. The results are obtained from multi-particle tracking and each bunch is composed of 3000 macro-particles.}
\label{fig:4.11}
\end{figure}

Simulation with macro-particles per bunch can be done similarly. Set 3000 macro-particles in each bunch and turn off the longitudinal coupled bunch instability in tracking, we give the results of bunch center shift and bunch length variation after 3000 turns in Fig.~\ref{fig:4.10}. The bunch profiles are given in Fig.~\ref{fig:4.11}. Compared with the results in  Fig.~\ref{fig:4.8} and Fig.~\ref{fig:4.9}, single-particle and multi-particle tracking show good agreements.   

The transient beam loading will lead to longitudinal coupled bunch instability once the natural damping is not strong enough. If so, the low-level RF feedback, RF feedforward, or longitudinal bunch-by-bunch feedback have to be applied to stabilize the beam. Here we briefly introduce the stabilization mechanism when a low-level RF feedback is applied. From the control theory, when the low-level RF feedback is included in the control loop, the impedance  sampled by the beam is modified by 
\begin{equation}\label{eq:4.3.3}
Z_{cl}(\omega) = \frac{Z(\omega)}{1+ \exp(-i\omega \tau)G(\omega)Z(\omega)\exp(i\phi)},
\end{equation}
where $\tau$ is the overall loop delay, $\phi$ is the loop phase adjusted and $G(\omega)$ is the gain. In Ref.~\cite{boussard1995beam}, it is shown that the minimum value of the impedance beam can sample is 
\begin{equation}\label{eq:4.3.4}
R_{min} = \frac{2}{\pi}\tau \frac{R_s}{Q}\omega_{rf}.
\end{equation}
For a rough estimation, in the harmonic cavity of PETRA-IV, with an overall loop delay $\tau=150$ ns, the $R_{min}$ would be decreased by a factor of 20.

\begin{figure}[htp]
\includegraphics[width=1\linewidth]{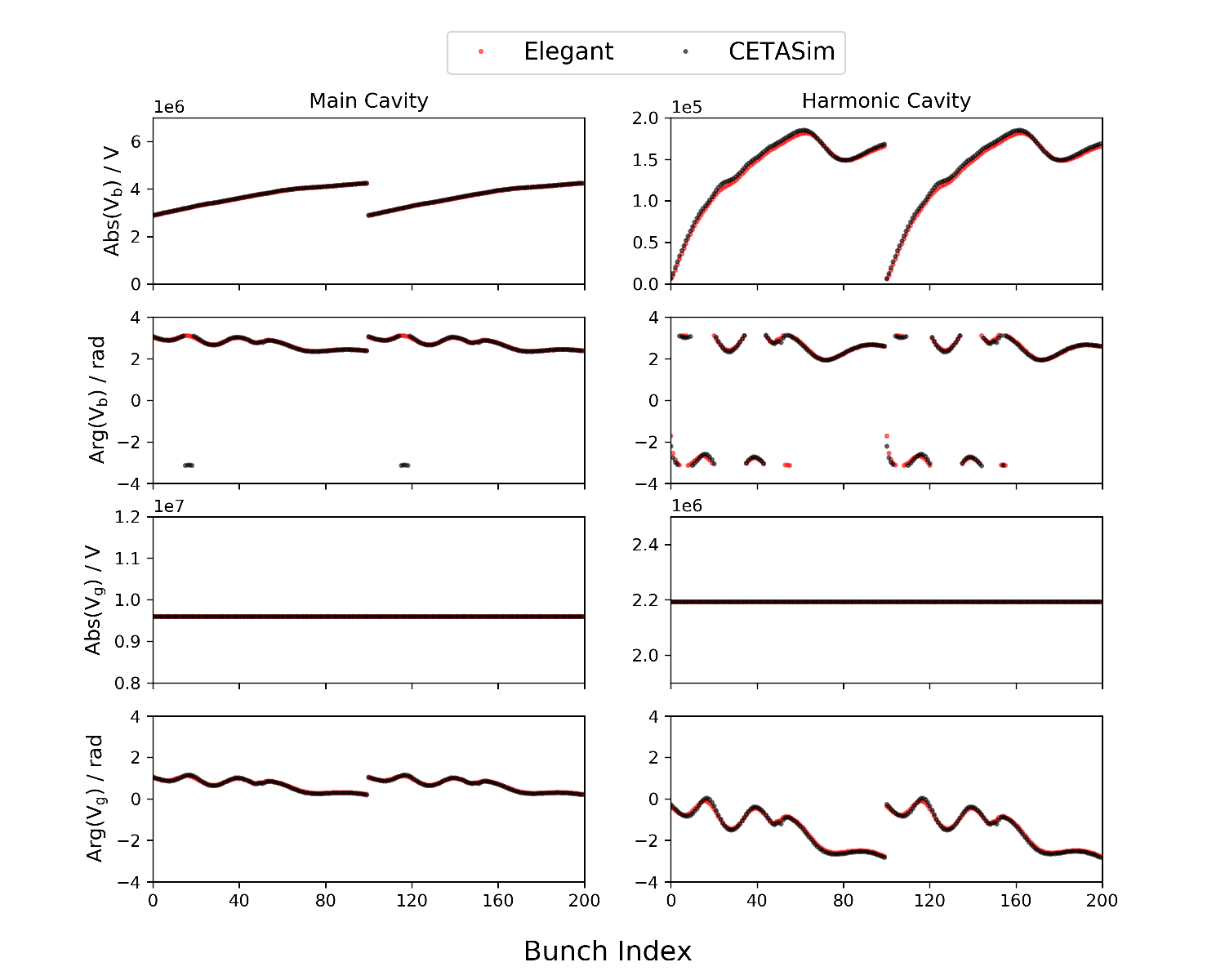} 
\caption{The amplitude and phase of the beam induced voltage $\boldsymbol{\tilde{V}_b}$ and generator voltage $\boldsymbol{\tilde{V}_g}$ each bunch sampled at the main cavity (left) and the harmonic cavity (right) after 1000 turn tracking by Elegant and CETASim.}
\label{fig:4.12}
\end{figure}

In the following, we show the benchmark between CETASim and Elegant of the transient beam loading simulation when the coupled bunch instability is turned on. To stabilize the beam, the shunt impedance of the main cavity and harmonic cavity are reduced by a factor of 2 and 20 respectively. Correspondingly, the cavity de-tuning frequencies $\Delta f_n$ are modified to $\Delta f_1=-13.953$ kHz and $\Delta f_3=13.882$ kHz to maintain the "optimized" de-tuning condition. The beam filling pattern is set as $h=3840=2\times(100\times(1+9)+920)$ which means there are 2 bunch trains and each bunch train includes 100 bunches. The total beam current is set to 200 $mA$ as well. Fig.~\ref{fig:4.12} shows the comparison of the beam-induced voltage and generator voltage on the bunch center at the 1000$th$ turn. Both in the main cavity and harmonic cavity, results from Elegant and CETASim agree well.

\subsection{4.5 Beam-ion effect}
As shown in Eq.~\ref{eq:3.6.4}, ions with a mass number larger than $A_{c}$ can be trapped in rings. If the oscillation of the trapped ions is confined within the beam pipe, beam-ion instability could be excited. Roughly speaking, with a high bunch charge, ions are over-focused and get lost within the gaps between bunches. With a low bunch charge, more ions can be trapped. However, if the bunch charge is too small, only a few ions can be ionized, leading to a weak beam-ion effect as well \cite{li2020beam}. In this sense, we expect the beam-ion effect to be of interest in the medium bunch charge region. In CETASim, the ion motion is limited in the transverse direction. CETASim supplies several parameters to act as comprise of simulation speed and accuracy, such as the number of macro-ions generated per collision, the transverse range beyond which the ions are cleaned, the number of beam-ion interactions per turn, etc. Below, we give two examples of the beam-ion effect simulation in PETRA-IV storage ring, single and multi-ion species. The filling pattern is set as the brightness mode operation scheme, $h=3840=80\times(20\times(1+1)+8)$, and the bunch charges are identical among the 1600 bunches. It is assumed that the total gas pressure is 1 $nTor$ and the gas temperature is 300 $K$. We set one beam-ion interaction point in one turn and take the value of the average betatron function to get the one-turn transfer matrix. In the simulation, one electron bunch is represented by one macro-electron particle. The electron bunch position is initialized on the axis. Ions are cleaned when their transverse distances reference to the ideal orbit are larger than 10 times the maximum effective beam size $abs(\boldsymbol{X_i})>10 \times (abs(\langle \boldsymbol x \rangle) + \boldsymbol{\sigma_{ x})}$. 

\begin{figure}[!htp]
\includegraphics[width=1\linewidth]{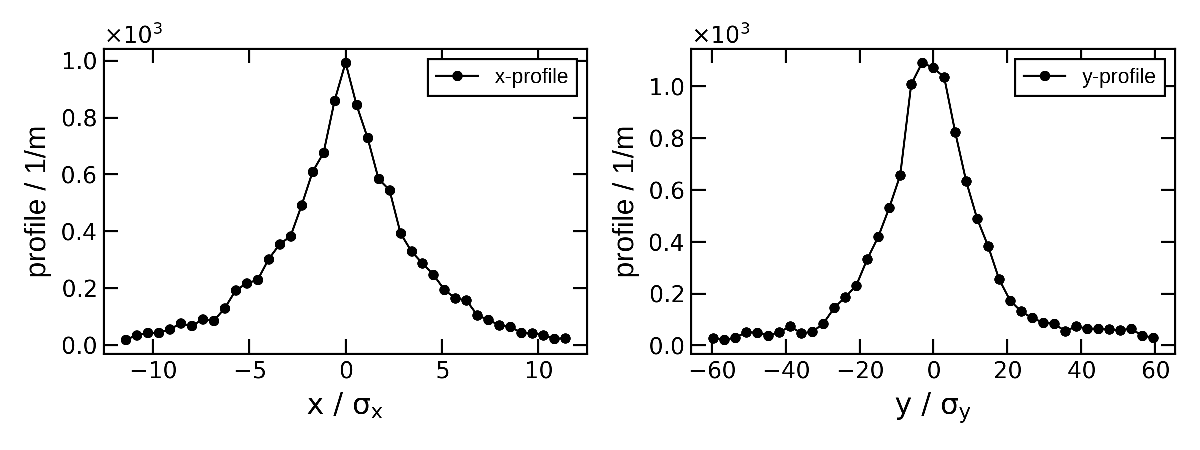} 
\caption{The equilibrium ionized $CO^{+}$ profile on x and y plane. The total beam current in simulation is set as 5 $mA$. The profiles of the accumulated ion cloud does not follow the Gaussian shape. The ion gas is composed of $CO$ only.}
\label{fig:4.5.1}
\end{figure}

\begin{figure}[!htp]
\includegraphics[width=1\linewidth]{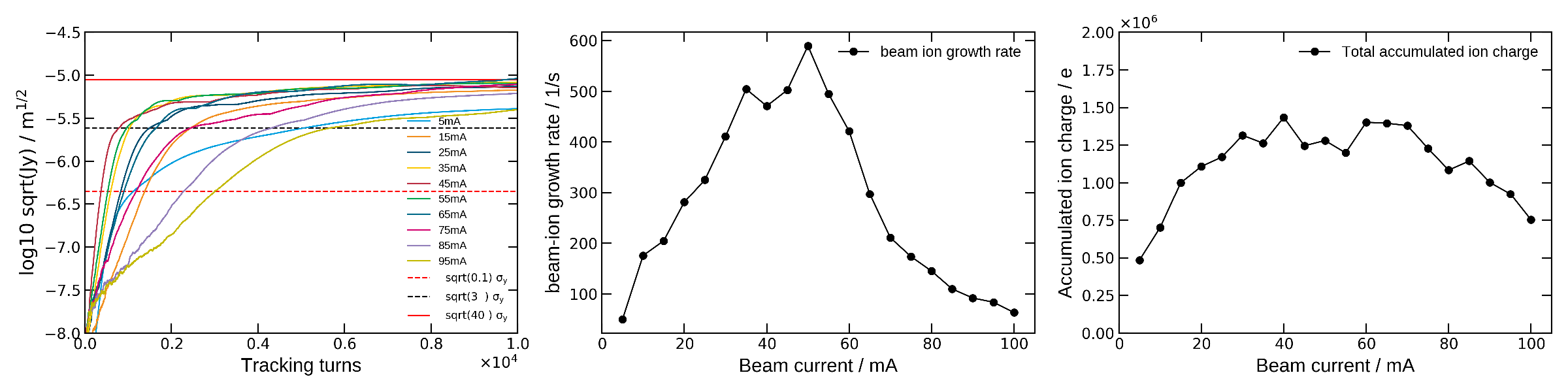} 
\caption{Left: the square root of the maximum vertical action $\sqrt{J_y}$ among 1600 bunches as a function of tracking turns; Middle: beam ion growth rate as a function of the total beam current; Right: the final accumulated ion charge after 10 K turns tracking as a function of the total beam current. The ion gas is composed of $CO$ only. The maximum bunch action $\sqrt{J_y}$ saturate to a value around $\sqrt{40 \epsilon_y}$.}
\label{fig:4.5.2}
\end{figure}

Set residual gas composed by $CO$ only, Fig.~\ref{fig:4.5.1} shows the equilibrium transverse ion $CO^{+}$ profile after 10 K turns tracking when the total beam current is 5 $mA$. For each beam-ion collision, 50 macro-ions are generated. The ion density profile does not follow a Gaussian shape as pointed out in Ref.~\cite{LanFaBeamIon}. Fig.~\ref{fig:4.5.2} (left) gives the square root of the maximum vertical action $\sqrt{J_{y}}$ among 1600 bunches as a function of tracking turns. The action $\sqrt{J_{y}}$ is defined as 
\begin{equation}\label{eq:4.4.1}
J_{y}   = \frac{1}{2} (\frac{1+\alpha_y^2}{\beta_y}y^2 + 2 \alpha_y y p_{y} + \beta_y p_{y}^2),
\end{equation}
where $\alpha_y$ and $\beta_y$ are the Twiss functions at the beam-ion interaction points. The square root of the bunch action $\sqrt{J_y}$ increases firstly and then gets saturated gradually to a value around $\sqrt{40 \epsilon_y}$. Fig.~\ref{fig:4.5.2} (middle) gives the growth rate of $\sqrt{J_{y}}$ as a function of the total beam current. The growth rate is obtained by an exponential fitting of $\sqrt{J_y}$ data selected in between ($\sqrt{0.1 \epsilon_y}$, $\sqrt{3\epsilon_y}$), indicating that the beam-ion effect is more severe in the median beam current region.  Figure~\ref{fig:4.5.2} (right) shows the final accumulated $CO^{+}$ charge as a function of the beam current. Again, at the median beam current region, more ions could be accumulated.

\begin{figure}[!htp]
\includegraphics[width=1\linewidth]{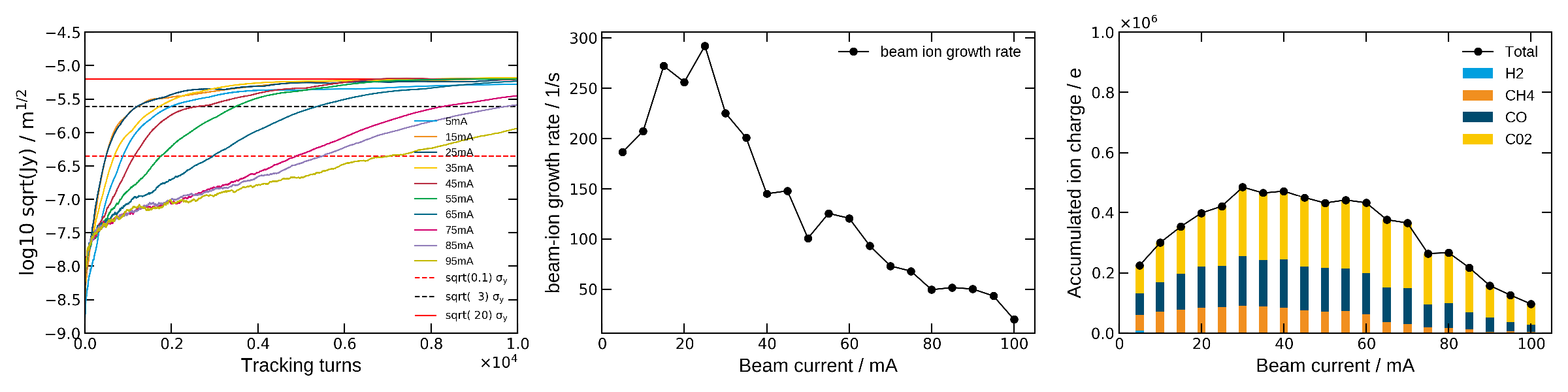} 
\caption{Left: the square root of the maximum vertical action $\sqrt{J_y}$ among all bunches as a function of tracking turns; Middle: beam ion growth rate; Right: the final accumulated ion charge after 10 $K$ turns tracking. The maximum bunch action $\sqrt{J_y}$ saturate to a value around $\sqrt{20 \epsilon_y}$.}
\label{fig:4.13}
\end{figure}

In the second simulation, the initial gas composition is set the same as that in APS-U \cite{Petra4CDR}. The residual gas is composed of $H_2$, $CH_4$, $CO$ and $CO_2$. The percentage of each gas is 0.43, 0.08, 0.36 and 0.13. Simulation results are shown in Fig.~\ref{fig:4.13}. Similar to the results obtained from the single-ion gas setting, the beam-ion effect is more severe in the median beam current region. The $\sqrt{J_y}$ saturates gradually to a value around $\sqrt{20 \epsilon_y}$. The growth rate is reduced roughly by a factor of 2. The sub-figure on the right gives the ion charge accumulated finally after 10 K tuns tracking. Clearly, the higher the total beam current is, the less of the lighter ions can be trapped. The total accumulated ion charge decreased roughly by a factor of 2 as well.

There are several things noticeable in the beam-ion study. Firstly, the beam-ion instability is usually self-limited. Secondly, according to the $\sqrt{J_y}$ selected for fitting, the growth rate will be different. Thirdly, the convergence of the simulations as different parameter settings, such as the number of beam-ion interaction points, the number of ions generated per beam-ion collision, the range to clean ions, etc, has to be studied beforehand. From our experience, cutting the ring into 10 sections usually is good enough to get a convergent ion charge accumulated in simulation. Set 10 macro-ions per collision is good comprising between simulation speed and accuracy. Set ion cleaning condition $m$ as 10  ($abs(\boldsymbol{X_i})>m \times (abs( \langle \boldsymbol  x \rangle) + \sigma_{\boldsymbol  x})$) is usually a good choice. The last thing is about the incoherent effect. If one wants to simulate the bunch emittance growth instead of the coherent bunch center oscillation, multi-electron particles can be set in the simulation as well. However, it is still the $Bassetti-Erskine$ model applied to get the Coulomb force among ions and electrons. In the future, a Poisson solver based on the Particle-In-Cell methods would be added to the beam-ion module to handle the problem of self-consistence in the 'strong-strong' simulation.

\subsection{4.6 Bunch-by-bunch feedback}
The bunch-by-bunch feedback samples the beam transverse or longitudinal centroid information in the bunch-by-bunch sense \cite{Lonza:1100539, nakamura2009single}. Passing these position signals through an FIR filter, the feedback creates bunch-by-bunch momentum kicks to the beam.  In Ref.~\cite{nakamura2009single}, Nakamura shows that the coefficients $a_k$ for an FIR filter can be found by the time domain least square-fitting (TDLSF) method. The oscillation of the beam at the $k$th turn is approximated by 
\begin{equation}\label{eq:4.5.1}
\begin{split}
x[k] &= \sum_{m=0}^{M} A^{m} \cos((1+\Delta_k^m)\phi_k^m + \psi^m) \\
&\approx \sum_{m=0}^{M} (P_0^m \cos(\phi_k^m) + P_1^m \phi_k^m \sin(\phi_k^m) + Q_0^m \sin(\phi_k^m) + Q_1^m \phi_k^m \sin(\phi_k^m)) 
\end{split}
\end{equation}
where $M$ is the number of oscillation of the beam, $A^m$ is the amplitude, $\phi_k^m$ and $(1+\Delta^m)\phi_k^m$ are the assumed and actual phase advance at the $k$th turn, $P$ and $Q$ are undefined parameters. In the same way, the output of the filter at current turn can be found  
\begin{equation}\label{eq:4.5.3}
\begin{split}
y[0] = \sum_{m=0}^{M} G^m &(P_0^m \cos(\varphi^m+\zeta^m) + P_1^m \varphi^m \sin(\varphi^m+\zeta^m) \\ & + Q_0 \sin(\varphi^m+\zeta^m) + Q_1^m \varphi^m \cos(\varphi^m+\zeta^m)) 
\end{split}
\end{equation}
where $G^m$ and $\zeta^m$ are the required gain and phase shift if the FIR filter, $\varphi^m$ is the assumed phase advance from BPM to the kicker. The FIR coefficients connects $x[k]$ to $y[0]$ with the required values of $G^m$, $\zeta^m$, $\varphi^m$ and $\phi_k^m$. Assuming the fitting function as $S=\sum_{k=0}^N(x[-k]-x_{-k})^2$, the least square fitting method would reduced to the condition
\begin{equation}\label{eq:4.5.4}
\begin{split}
\frac{\partial S}{\partial P_i^m } = \frac{\partial S}{\partial Q_i^m } = 0 
\end{split},
\end{equation}
by which the FIR coefficient $a_k$ can be obtained. 

\begin{figure}[!htp]
\includegraphics[width=1\linewidth]{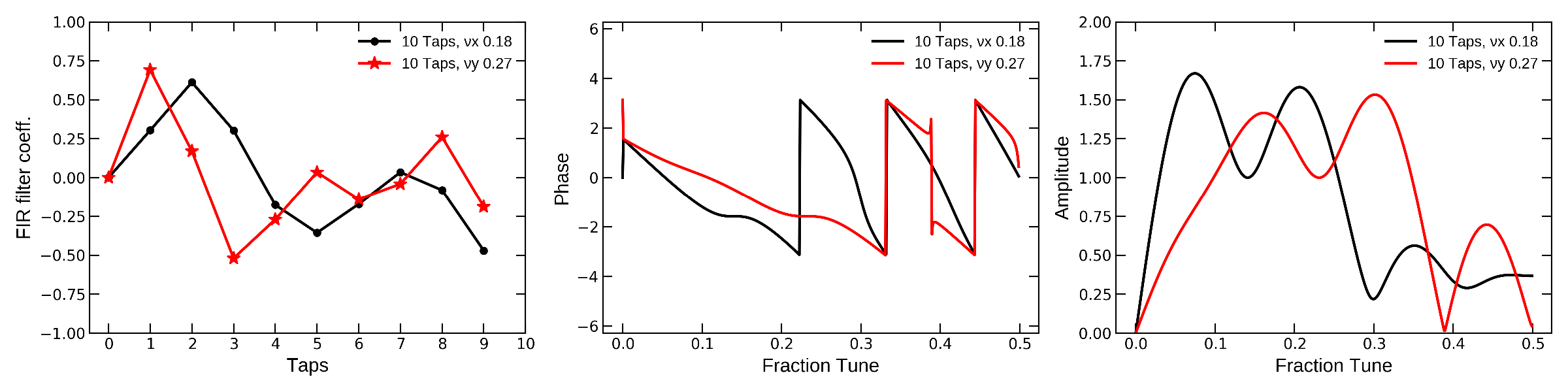} 
\caption{The horizontal and vertical 10-tap FIR filter coefficients $a_k$ (left), the frequency response of phase (middle) and amplitude (right). The horizontal and vertical target tune fractions are 0.18 and 0.27.}
\label{fig:4.14}
\end{figure}

Following this TDLSF method, a preliminary 10-tap FIR filter is designed, in the transverse directions for the PETRA-IV storage ring. The FIR filter has a zero amplitude response at the frequency $n f_0$ (DC rejection), so that the components caused by the closed orbit distortions, unequal bunch signal shapes from pickup electrodes, reflection at cable connections, $etc$, are filtered out. The information of the current turn is dropped off, indicating a one-turn delay $a_{0}=0$. The first derivative of the phase response curves at the fraction of the target betatron tune $\nu f_0$ is designed to be zero to enlarge the phase error tolerance. The normalized amplitude response at the target tune fraction is a local minimum. The stable working region is limited by the filter phase response curve within $(-\pi,0)$.  Fig.~\ref{fig:4.14} shows the filter coefficients, the phase and amplitude responses as functions of the tune fractions. In the simulation study, the pickup and kicker are located at the same place which means the phase responses of the filter at the target tunes have to be $-\pi/2$. If the maximum power $P_{max}$ and the kicker impedance $Z_{kicker}$ are specified in the input file, the kicker voltage will be limited to $\sqrt{P_{max} Z_{kick}}$.   

\begin{figure}[!htp]
\includegraphics[width=1\linewidth]{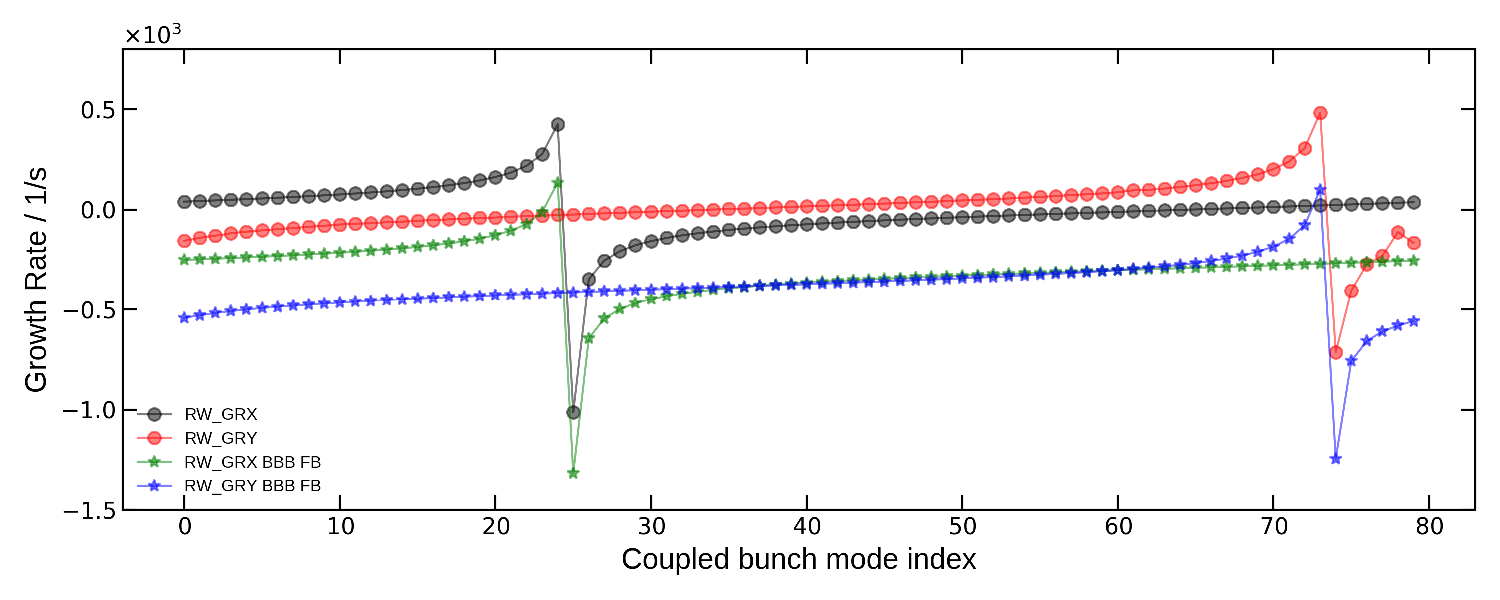} 
\caption{Comparison of the transverse coupled bunch growth rate due to the RW  with and without bunch-by-bunch feedback. The simulation condition is the same as those applied in Fig.~\ref{fig:4.5}. The 10 taps bunch-by-bunch feedback shown in Fig.~\ref{fig:4.14} is applied in tracking. }
\label{fig:4.15}
\end{figure}

\begin{figure}[!htp]
\includegraphics[width=1\linewidth]{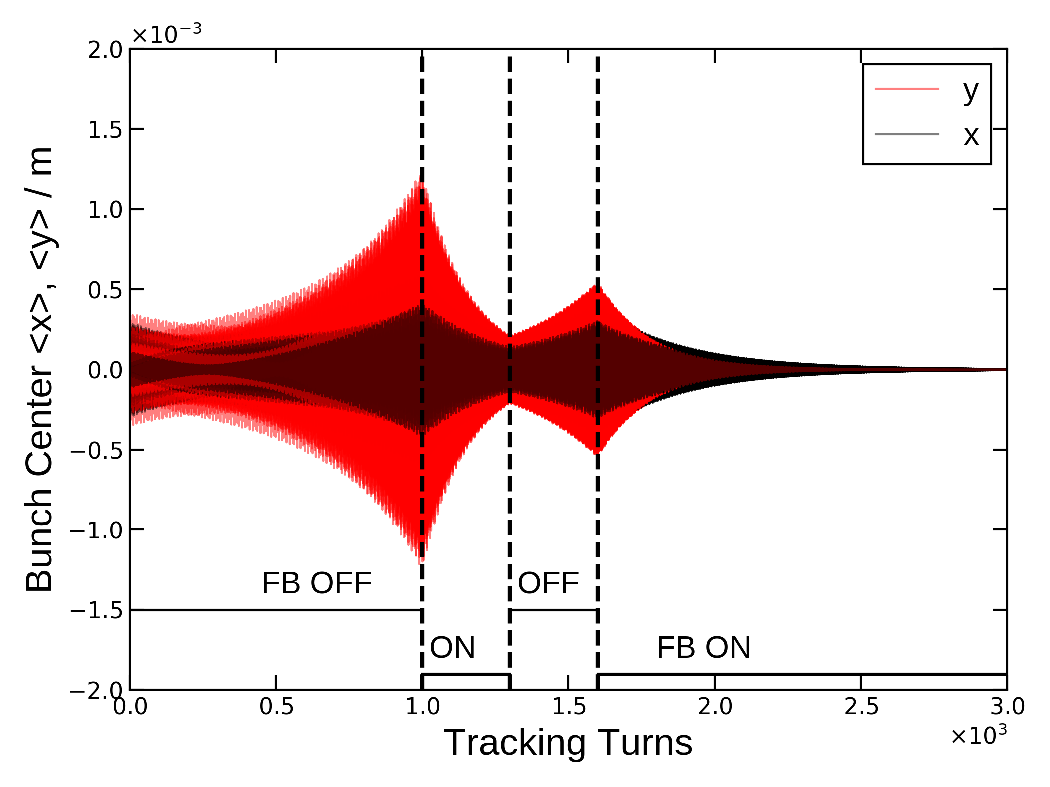} 
\caption{Trajectories of the 80 bunches centroids. The bunch-by-bunch feedback is turned on from 1000 to 1300 turns and from 1600 to 3000 turns during tracking. The long-range RW wakes are turned on during the whole simulation.}
\label{fig:4.16}
\end{figure}

Figure~\ref{fig:4.15} shows the comparison of the transverse coupled bunch mode growth rate due to the RW wakes with and without bunch-by-bunch feedback. The simulation conditions are the same as those applied in Fig.~\ref{fig:4.5}. The 10 taps FIR filter bunch-by-bunch feedback is also applied in tracking. All modes are suppressed. With the same beam condition, we also give the motions of the 80 bunch centroids in the "grow-damped" simulation in Fig.~\ref{fig:4.16}. During the tracking, the bunch-by-bunch feedback is turned on from 1000 to 1300 turns and from 1600 to 3000 turns. Clearly, without the feedback, the bunches are unstable due to the transverse long-range RW wakes. When the bunch-by-bunch feedback is turned on, oscillations of all of the bunches can be stabilized to zero. 

\subsection{4.7 Emittance exchange and linear coupling}
Skew quadrupole leads to a coupling effect between the horizontal and vertical planes, which re-distributes the equilibrium emittances together with the synchrotron radiation damping and quantum-excitation effects. Here, we note $K$ as the skew quadruple strength with dimension $1/m$. In Ref.~\cite{RyanLindberg}, Lindberg shows a model to predict the equilibrium emittance as a function of the difference of fraction tune $\Delta_r$,  
\begin{equation}\label{eq:4.6.1}
\begin{split}
\epsilon_x &= \epsilon_0 \frac{1+\frac{1}{4 \tau_x}(\tau_y-3\tau_x) \sin^2\theta }{1+\frac{1}{4 \tau_x \tau_y}(\tau_x-\tau_y)^2 \sin^2\theta}  \quad  \quad 
\epsilon_y = \epsilon_0 \frac{\frac{1}{4 \tau_x}(\tau_y + \tau_x) \sin^2\theta }{1+\frac{1}{4 \tau_x \tau_y}(\tau_x-\tau_y)^2 \sin^2\theta}
\end{split}
\end{equation}
where, $\sin^2\theta=\frac{\kappa^2}{\kappa^2+\Delta_r^2}$ and  $\kappa = \sqrt{\frac{\beta_x \beta_y}{4 \pi^2} K^2}$ is the linear coupling coefficient. If the working tune smoothly crosses the difference resonance, the emittance would be exchanged between the horizontal and vertical planes due to this linear coupling.

\begin{figure}
\centering
\subfigure{
    \label{fig4.17_a} 
    \includegraphics[width=0.45\textwidth]{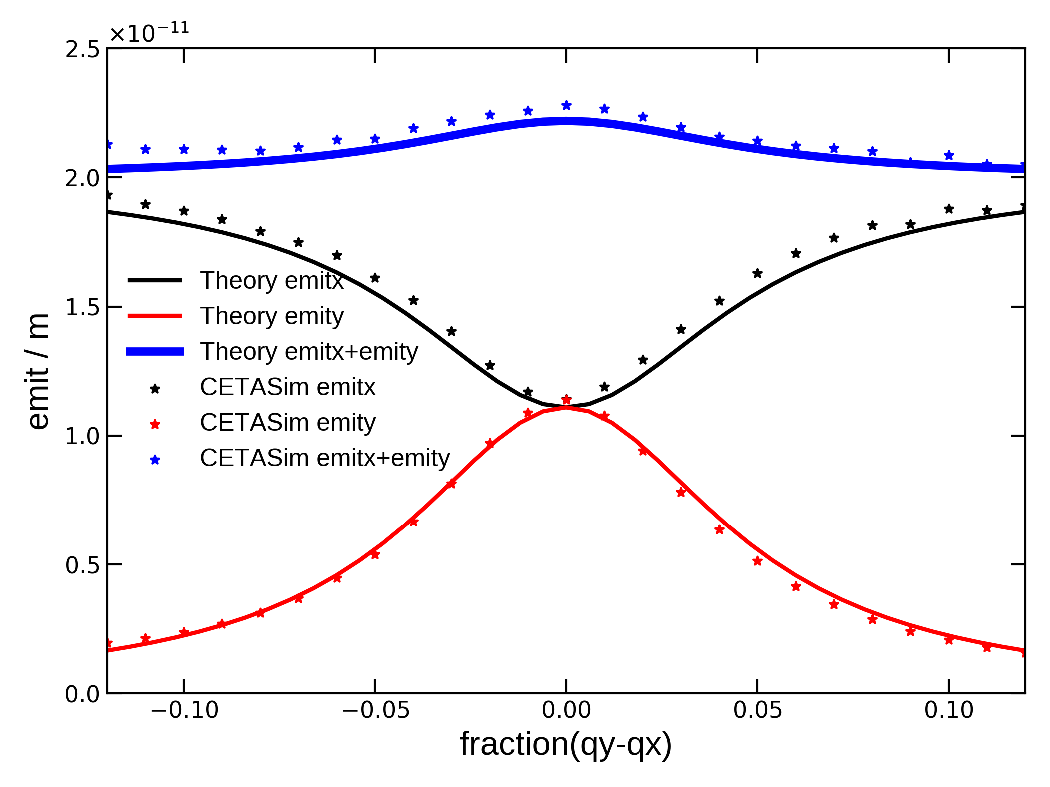}}
\subfigure{
    \label{fig4.17_b} 
    \includegraphics[width=0.45\textwidth]{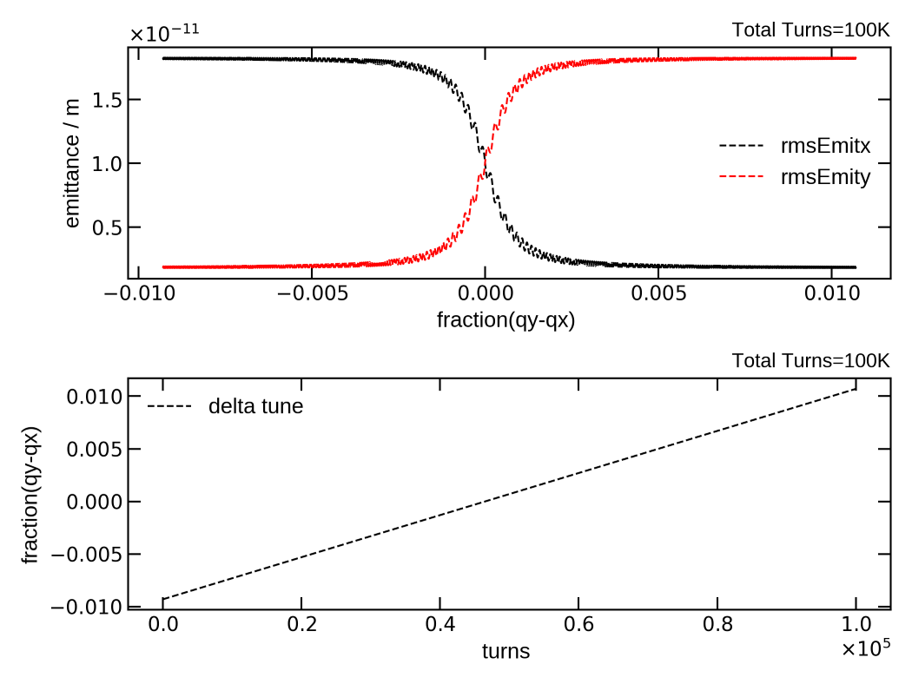}} 
\caption{\label{fig:4.17}Left: equilibrium emittance as a function of distance to difference resonance. Right: The horizontal and vertical emittance are exchanged when the difference resonance is smoothly crossed in $10^5$ turns.}
\end{figure}

To study the effect, an extra skew quadrupole can be set beside the one-turn transfer map Eq.~\ref{eq:3.1.2} in CETASim. Both the strength of the skew quadrupole and the working tunes can be set as ramping variables as a function of the tracking turns. We give two cases for linear coupling studies. The first one is the static case in which all the lattice parameters are fixed during tracking. The second one is the dynamical cases, during which the difference resonance is crossed by ramping the working tune. Still, the nominal settings of the PETRA-IV lattice and beam condition are applied as the initial conditions. Fig.~\ref{fig:4.17} shows the simulation results from CETASim. The left sub-figure shows the final equilibrium emittance as a function of tune difference in the static simulation. The skew quadrupole strength is set as 0.05 $1/m$. Compared to the predictions from Eq.~\ref{eq:4.6.1}, tracking results show good agreements.  The sub-figure on the right shows how the horizontal and vertical emittance are exchanged when the difference resonance is smoothly crossed. The skew quadrupole strength is set to $5 \times 10^{-4}$ $1/m$, and the vertical tune ramps from 0.17 to 0.19 within $10^5$ turns.  

\section{5 summary and outlook}
In the report, the code CETASim, developed recently, is introduced in detail. The motivation for CETASim development is to have a simulation tool that covers the collective effects in electron storage rings, especially when different filling pattern schemes are of great concern. The charge of the bunches can be set differently to study the effects of the “guarding bunches” which are normally used for transient beam loading compensation and ion cleaning. The architecture of the code is carefully designed so that one can expand the code without too many difficulties if some other beam dynamics have to be taken into account.  Instead of the element-by-element tracking method, CETASim takes the one-turn transfer map for simplicity, where the amplitude-dependent tune shift and the momentum compaction factor can be taken into account up to the second order. Currently,  CETASim includes modules to study the single-bunch effects, coupled-bunch effects, transient beam loading, beam-ion effects and bunch-by-bunch feedback. These modules are benchmarked either with results from the theoretical predictions or tracking from Elegant. For the study of the transient beam loading, the coupled generator dynamics and beam dynamics are treated self-consistently. The coupled-bunch instability can be turned off, which is not physical, however, it can help the user to get the first idea about how different bunches are lengthened due to different filling patterns.

Still, there are several things to be upgraded in the future. The first one is to have a subroutine that can import the external short-range and long-range wakes and apply them in simulation. As we mentioned in the report, for the single bunch effect, CETASim takes the impedance as the green function; and for the coupled bunch effect, the long-range wakes are limited to the analytical RW and RCL models. The second one is to have a module to simulate the cavity feedback self-consistently. At this moment, a very simple ideal cavity feedback is available. In the future, we will update the cavity feedback module to cover the real experiment setups with the help of the low-level RF group. The third one is the Coulomb interaction between electrons and ions.  The $Bassetti-Erskine$ formula based on the 2D Gaussian distribution function is not an accurate approach to get the Coulomb force among the ions and electrons. A self-consistent PIC subroutine will be developed to handle this problem. Finally, we plan to improve the performance of CETASim including algorithm optimization and parallelization in the future.

\section{6 Acknowledgements}
The authors would like to thank Dr. Mikhail Zobov from INFN in Frascati for his helpful discussions. We would like to thank Dr. Sergey Antipov as well for proofreading the manuscript and providing numerous helpful comments and suggestions. This work is supported by the European Union’s Horizon 2020 research and innovation program under grant agreement No. 871072.

\bibliography{note}
\end{document}